%Include: vrtx1.eps, vrtx2.eps, vrtx3.eps, source.eps, sgraph1.eps,
%sgraph2.eps

\input harvmac
\input amssym.def
\input amssym.tex
\input epsf.tex
\def\tilde{\widetilde}
\def\bar{\overline}
\def\rk#1{\mathop{\rm rank}\nolimits(#1)}
\def\^{{\wedge}}
\def\wt{\widetilde}
\def\hat{\widehat}
\def\bar{\overline}
\def\BC{{\Bbb C}}

\def\BR{{\Bbb R}}
\def\bigwedge{\wedge}
\def\CL{{\cal L}}
\def\CM{{\cal M}}
\def\CN{{\cal N}}

\noblackbox

\def\urlfont{\hyphenpenalty=10000 \hyphenchar\tentt='057 \tt}

\newbox\tmpbox\setbox\tmpbox\hbox{\abstractfont PUPT-2128}
\Title{\vbox{\baselineskip12pt\hbox{\hss hep-th/0409149}
\hbox{PUPT-2128}}}
{\vbox{
\centerline{New Instanton Effects in Supersymmetric QCD}}}
\smallskip
\centerline{Chris Beasley}
\smallskip
\centerline{\it{Joseph Henry Laboratories, Princeton University}}
\centerline{\it{Princeton, New Jersey 08544}}
\medskip
\centerline{and}
\medskip
\centerline{Edward Witten}
\smallskip
\centerline{\it{School of Natural Sciences, Institute for Advanced
Studies}}
\centerline{\it{Princeton, New Jersey 08540}}
\bigskip\bigskip

In supersymmetric QCD with $SU(N_c)$ gauge group and $N_f$
flavors, it is known that instantons generate a superpotential if
$N_f=N_c-1$ and deform the moduli space of vacua if $N_f=N_c$. But
the role of instantons has been unclear for $N_f>N_c$. In this
paper, we demonstrate that for $N_f>N_c$, on the moduli space of
vacua, instantons generate a more subtle chiral operator
containing (for example) non-derivative interactions of
$2(N_f-N_c)+4$ fermions. Upon giving masses to some flavors, one
can integrate out some fermions and recover the standard results
for $N_f=N_c$ and $N_f=N_c-1$. For $N_f=N_c$, our analysis gives,
in a sense, a more systematic way to demonstrate that instantons
deform the complex structure of the moduli space of vacua.

\Date{September 2004}

\lref\AffleckMK{ I.~Affleck, M.~Dine and N.~Seiberg, ``Dynamical
Supersymmetry Breaking In Supersymmetric QCD,'' Nucl.\ Phys.\ B
{\bf 241} (1984) 493--534, ``Dynamical Supersymmetry Breaking In
Four-Dimensions and Its Phenomenological Implications,'' Nucl.
Phys. {\bf B256} (1985) 557--599.}

\lref\AmatiFT{
D.~Amati, K.~Konishi, Y.~Meurice, G.~C.~Rossi and G.~Veneziano,
``Nonperturbative Aspects In Supersymmetric Gauge Theories,''
Phys.\ Rept.\  {\bf 162} (1988) 169--248.}

\lref\AntoniadisHG{
I.~Antoniadis, E.~Gava, K.~S.~Narain and T.~R.~Taylor,
``Effective Mu Term in Superstring Theory,''
Nucl.\ Phys.\ B {\bf 432} (1994) 187--204,
{\urlfont hep-th/9405024}.}

\lref\AntoniadisQG{
I.~Antoniadis, E.~Gava, K.~S.~Narain and T.~R.~Taylor,
``Topological Amplitudes in Heterotic Superstring Theory,''
Nucl.\ Phys.\ B {\bf 476} (1996) 133--174,
{\urlfont hep-th/9604077}.}

\lref\BershadskyCX{
M.~Bershadsky, S.~Cecotti, H.~Ooguri and C.~Vafa,
``Kodaira-Spencer Theory of Gravity and Exact Results for Quantum
String Amplitudes,'' Commun.\ Math.\ Phys.\  {\bf 165} (1994)
311--428, {\urlfont hep-th/9309140}.}

\lref\BeasleyFX{
C.~Beasley and E.~Witten,
``Residues and World-Sheet Instantons,''
JHEP {\bf 0310} (2003) 065, {\urlfont hep-th/0304115}.}

\lref\CordesUM{
S.~F.~Cordes,
``The Instanton Induced Superpotential In Supersymmetric QCD,''
Nucl.\ Phys.\ B {\bf 273} (1986) 629--648.}

\lref\dsww{M. Dine, N. Seiberg, X.~G. Wen, and E. Witten,
``Nonperturbative Effects on the String Worldsheet, I, II,''
Nucl. Phys. {\bf B278} (1986) 769--789, Nucl. Phys. {\bf B289} (1987)
319--363.}

\lref\FinnellDR{
D.~Finnell and P.~Pouliot,
``Instanton Calculations Versus Exact Results in Four-Dimensional SUSY
Gauge Theories,'' Nucl.\ Phys.\ B {\bf 453} (1995) 225--239, {\urlfont
hep-th/9503115}.}

\lref\GatesNR{
S.~J.~Gates, M.~T.~Grisaru, M.~Rocek and W.~Siegel,
{\it Superspace, Or One Thousand And One Lessons In Supersymmetry},
Front.\ Phys.\ {\bf 58}, Benjamin/Cummings Publishing Company, Inc.,
Reading, Massachusetts, 1983,
{\urlfont hep-th/0108200}.}

\lref\Griffiths{
P.~Griffiths and J.~Harris, {\it Principles of Algebraic Geometry},
John Wiley and Sons, New York, 1978.}

\lref\IntriligatorAU{
K.~A.~Intriligator and N.~Seiberg,
``Lectures on Supersymmetric Gauge Theories and Electric-Magnetic
Duality,'' Nucl.\ Phys.\ Proc.\ Suppl.\  {\bf 45BC} (1996) 1--28,
{\urlfont hep-th/9509066}.}

\lref\LercheUY{
W.~Lerche, C.~Vafa and N.~P.~Warner,
``Chiral Rings In N=2 Superconformal Theories,''
Nucl.\ Phys.\ B {\bf 324} (1989) 427--474.}

\lref\Liu{K. Liu, ``Holomorphic Equivariant Cohomology,''
Math. Ann. {\bf 303} (1995) 125--148.}

\lref\KaplunovskyFG{
V.~Kaplunovsky and J.~Louis,
``Field Dependent Gauge Couplings in Locally Supersymmetric Effective Quantum
Field Theories,'' Nucl.\ Phys.\ B {\bf 422} (1994) 57--124,
{\urlfont hep-th/9402005}.}

\lref\SeibergBZ{
N.~Seiberg,
``Exact Results on the Space of Vacua of Four-Dimensional SUSY Gauge
Theories,'' Phys.\ Rev.\ D {\bf 49} (1994) 6857--6863,
{\urlfont hep-th/9402044}.}

\lref\SeibergPQ{
N.~Seiberg,
``Electric - Magnetic Duality in Supersymmetric Nonabelian Gauge Theories,''
Nucl.\ Phys.\ B {\bf 435} (1995) 129--146,
{\urlfont hep-th/9411149}.}

\lref\Shifman{ V. A. Novikov, M. A. Shifman, A. I. Vainshtein, and
V. I. Zakharov, ``Supersymmetric Instanton Calculus (Gauge
Theories With Matter),'' Nucl. Phys. {\bf B260} (1985) 157--181.}

\lref\ShifmanN{ M. A. Shifman, ``Exact Results In Gauge Theories:
Putting Supersymmetry To Work,'' Int. J. Mod. Phys. {\bf A14}
(1999) 5017--5048, {\urlfont hep-th/9906049}.}

\lref\Warner{
N.~P.~Warner, ``Lectures on N=2 Superconformal Theories and
Singularity Theory,'' in {\it Superstrings '89: Proceedings of the
Trieste Spring School}, pp. 197--237, Ed. by M.~Green et al, World
Scientific, Singapore, 1990.}

\lref\WittenU{E. Witten, ``An $SU(2)$ Anomaly,'' Phys.\ Lett.\ B {\bf
117} (1982) 324--328.}

\lref\WittenZZ{
E.~Witten,
``Mirror Manifolds and Topological Field Theory,'' in
{\it Essays on Mirror Manifolds}, pp. 121--160, Ed. by  S.~T.~Yau,
International Press Co., Hong Kong, 1992,
{\urlfont hep-th/9112056}.}

\newsec{Introduction}

Supersymmetric QCD (or SQCD) with gauge group $SU(N_c)$ and $N_f$
massless flavors has been extensively studied. (By a flavor, we
mean a massless chiral multiplet transforming in the fundamental
plus antifundamental representation.)  In particular, partly
because of holomorphy and the extensive symmetry group this theory
possesses, many properties of its low energy vacuum structure are
amenable to exact analysis. Yet the theory still displays a wealth
of interesting non-perturbative phenomena, including generation of
a superpotential
\refs{\AffleckMK\AmatiFT\Shifman\CordesUM{--}\FinnellDR},
deformation of the complex structure of the moduli space and
appearance of composite massless particles \refs{\SeibergBZ} (see
also \KaplunovskyFG\ for a related analysis), and
electric-magnetic duality \refs{\SeibergPQ}. For some short
reviews of some aspects of this story, see
\refs{\IntriligatorAU,\ShifmanN}.

The properties of the  theory depend very much on $N_f$. For
$N_f=0$, it is believed that a discrete chiral symmetry is
dynamically broken in this theory. The behavior for $N_f<N_c-1$
can largely be deduced from this statement \AffleckMK. For
$N_f=N_c-1$, instantons generate a superpotential which lifts all
flat directions on the moduli space ${\cal M}$ of supersymmetric
vacua. For $N_f=N_c$,   instantons do not generate a
superpotential, but rather \refs{\SeibergBZ} deform  the complex
structure of the moduli space ${\cal M}$.  For $N_f>N_c$, no
superpotential is generated and ${\cal M}$ is undeformed; the
important dynamical statements for $N_f>N_c$ concern the behavior
at the origin of ${\cal M}$, where composite massless particles
appear \refs{\SeibergBZ}, including \SeibergPQ\ gauge fields
related to electric-magnetic duality  if $N_f\geq N_c+2$.

Though the important statements for $N_f>N_c$ involve the behavior
at the origin in ${\cal M}$, it is also possible to explore the
behavior far from the origin for this range of $N_f$.  In that
region, the theory has gauge instantons whose effects are
computable, just as they are for $N_f=N_c-1$ and $N_f=N_c$. One
might well ask, ``What are these instantons good for?'' The
purpose of our paper is to answer this question. To keep things
simple, we concentrate on the case $N_c=2$, but we believe that
the story is similar for all $N_c$.

Our answer is that instantons for all $N_f\geq N_c-1$ generate a
chiral interaction, or $F$-term, on ${\cal M}$.  For $N_f=N_c-1$,
this is the familiar instanton-induced superpotential.  For
$N_f=N_c$, the $F$-term that is generated is a four-fermion (or
two-derivative) interaction on ${\cal M}$ which describes the
familiar deformation of the complex structure of ${\cal M}$. In
fact, understanding the complex structure deformation in this
slightly novel way gives, possibly, a more systematic way to
understand how it comes about, and which ${\cal N}=1$ theories
have such a deformation. For $N_f>N_c$, we get more exotic
$F$-terms on ${\cal M}$ which generate, for example, interactions
with $2(N_f-N_c)+4$ fermions.  These interactions produce no
obvious qualitative effect in the physics as long as the $N_f$
flavors are massless, but if one adds bare masses for some
flavors, they induce the usual instanton effects for $N_f=N_c$ and
$N_f=N_c-1$.

$F$-terms of the type that appear in our analysis -- we will call
them multi-fermion $F$-terms -- have been little discussed in
${\cal N}=1$ supersymmetric theories in four dimensions.  This
partly accounts for the fact that their appearance in SQCD has not
been noted until now. Interactions of this type are far more
familiar in the closely related context of two-dimensional  sigma
models with ${\cal N}=(2,2)$ supersymmetry (which can arise by
compactification or dimensional reduction from four dimensions);
in that context, the multi-fermion $F$-terms are associated with
the generators of the $(c,c)$ chiral ring, or equivalently the
ring of local observables of the topological $B$-model.

We begin our analysis in section 2 with a general description of
multi-fermion $F$-terms in $\CN=1$ supersymmetric effective
actions. In section 3, we analyze, for $N_c=2$, the relevant terms
that are possible in the specific case of supersymmetric QCD.  We
show that, for each value of $N_f$, their structure is uniquely
determined by the flavor symmmetries. Here we exploit the fact
that for $N_c=2$, the flavor symmetry is enhanced because both the
quarks and the anti-quarks transform in the fundamental
representation of the gauge group.

Finally, we show in section 4 that these multi-fermion $F$-terms
are indeed generated in the low energy effective action of SQCD.
We do this in three ways, each of which casts a different light on
the origin of these unusual corrections. First, we perform a direct
instanton computation as in \AffleckMK\ to show that the multi-fermion
$F$-terms are generated.  Second, in the special case that $N_c=2$,
$N_f =N_c+1= 3$, we show that they arise from a tree-level Feynman diagram
computation in the Seiberg dual description of the theory.  Third, we
show that upon perturbing the theory by supersymmetric bare masses for some
flavors, these interactions give rise by renormalization group
flow to the standard results for $N_f=N_c$ and $N_f=N_c-1$.  This
result generalizes to $N_f>N_c$ the standard renormalization group
flow \SeibergBZ\ from the instanton-induced deformation of moduli
space for $N_f=N_c$ to the instanton-induced superpotential for
$N_f=N_c-1$.  We expect that these analyses will have analogs for all
$N_c$ and $N_f$.

Although our focus in this paper is on four-dimensional gauge
theory, multi-fermion $F$-terms also appear naturally in
supersymmetric effective actions that describe four-dimensional
compactifications of string theory. In fact, the present work
began with our asking whether worldsheet instantons of the
heterotic string can deform the complex structure of the moduli
space of supersymmetric vacua, like gauge instantons in SQCD for
$N_f=N_c$.  The question is motivated by the fact that \dsww\ such
worldsheet instantons can sometimes generate a spacetime
superpotential, like gauge instantons for $N_f=N_c-1$. The answer
to the question is positive, as we will explain, along with other
results, in a separate paper. A closely related matter is the
analysis \refs{\AntoniadisHG,\AntoniadisQG} of holomorphic
anomalies in the low energy effective action of the heterotic
string.

\newsec{General Remarks on Multi-Fermion $F$-Terms}

In this section, we describe the general structure of $F$-terms
containing many fermions in $\CN=1$ supersymmetric effective
actions. However, before discussing generalities, we will motivate
our study of these interactions by considering a very specific and
well-known example: the complex structure deformation of the
moduli space ${\cal M}$ of vacua that occurs in $SU(2)$ SQCD with
two flavors.

When the distinction is important, we write ${\cal M}_{cl}$ for
the classical approximation to the moduli space of supersymmetric
vacua, and ${\cal M}$ for the exact quantum moduli space.

\subsec{Example: $SU(2)$ SQCD With Four Doublets}

For $N_c=2$, that is for gauge group $SU(2)$, the fundamental and
antifundamental representations coincide, so the $SU(2)$ gauge
theory with $N_f$ flavors is more naturally described as a theory
with $2N_f$ doublets, that is, with chiral multiplets transforming
as $2N_f$ copies of the two-dimensional representation ${\bf 2}$.
Since the term ``flavors'' is a little misleading for $SU(2)$, we
will write the number of doublets as $2n$.  (In $SU(2)$ gauge
theory with chiral multiplets transforming as doublets, the number
of such doublets must be even because of a global anomaly
\WittenU.)

To establish notation for the rest of the paper, we combine the matter
fields into one chiral multiplet,
\eqn\QS{ Q^i_a \,=\, q^i_a \,+\, \theta \psi^i_a \,+\, \cdots,}
with $a=1,2$ being the color index for the {\bf 2} of the $SU(2)$
gauge symmetry, and $i=1,\ldots,2n$ being the flavor index for the
{\bf 2n} of the global $SU(2n)$ flavor symmetry.  Of course, we
have indicated in \QS\ the component expansion of $Q^i_a$,
including a scalar field $q^i_a$ and a Weyl fermion $\psi^i_a$.

We also introduce the gauge invariant, composite meson chiral
superfield $M^{i j}$, given by \eqn\MS{ M^{ij} = \epsilon^{ab} \,
Q^i_a Q^j_b.} The meson $M^{i j}$ is clearly anti-symmetric in the
flavor indices $i$ and $j$ and so transforms in the skew
representation $\bigwedge^2({\bf 2n})$ of $SU(2n)$.

Using the mesons $M^{i j}$, we can succinctly describe the
classical moduli space ${\cal M}_{cl}$ of supersymmetric vacua as
being parametrized by arbitrary expectation values of $M^{i j}$
subject to the constraint \eqn\PLUCKER{ M \^ M \,=\, 0\,,} or more
explicitly, \eqn\PLUCKERII{ \epsilon_{i_1 j_1 i_2 j_2 \cdots i_n
j_n} \, M^{i_1 j_1} M^{i_2 j_2} \,=\, 0\,.} This system of
quadratic equations \PLUCKER\ simply enforces the condition that
\eqn\RKM{ \rk{M} \le 2\,,} as follows from the definition \MS\ of
$M^{i j}$ as the skew product of two quark superfields.

Now, if the number of doublets is $2n=4$, the classical constraint
\PLUCKER\ reduces to a single quadratic equation
\eqn\quadeq{\epsilon_{i_1j_1i_2j_2}M^{i_1j_1}M^{i_2j_2}=0} which
must be satisfied by $M^{i j}$.  Upon introducing suitable complex
linear combinations $m^I$, ${I = 1,\ldots,6}$, of the six
independent components $M^{ij}$, $i,j=1,\dots,4$, so as to
diagonalize the nondegenerate quadratic form that appears on the
left hand side of \quadeq, the classical equation \quadeq\ becomes
\eqn\PLUCKERIII{ \sum_{I=1}^6 \, \left(m^I\right)^2 \,=\, 0\,.}
The classical moduli space ${\cal M}_{cl}$ is thus smooth away
from the origin.  Its singularity at the origin is a signal of the
unbroken gauge symmetry. The $m^I$ transform in the vector
representation of the $SU(4)$ or $SO(6)$ flavor symmetry of the
$SU(2)$ gauge theory with four doublets.

The classical moduli space ${\cal M}_{cl}$ whose structure we have
just reviewed is deformed in the quantum theory \refs{\SeibergBZ}
and does not coincide with the quantum moduli space of vacua
${\cal M}$. To describe this deformation, we introduce the usual
holomorphic coupling scale $\Lambda$. Then, in the quantum theory,
the moduli space ${\cal M}$ is described by the modified
constraint \eqn\QPLUCKER{ M \^ M \,=\, \Lambda^4\,,} or
equivalently, with $\epsilon \sim \Lambda^4$, \eqn\QPLUCKERII{
\sum_{I=1}^6 \, \left(m^I\right)^2 \,=\, \epsilon\,.} Up to a
multiplicative constant, the form of the deformation \QPLUCKER\ is
determined completely by the $SU(4)$ flavor symmetry and
dimensional analysis.  Of course, as a result of the deformation,
the singularity of ${\cal M}_{cl}$ at the origin is removed and
${\cal M}$ is a smooth complex manifold.

\medskip\noindent{\it Representing the Deformation in the Effective
Action}\smallskip

But precisely how does the deformation \QPLUCKER\  appear in the
effective action of SQCD?  Is there a term of order $\epsilon$ in
the  low energy effective action whose presence signals this
deformation?

In this very simple example, one way to implement the quantum
deformation in the low energy effective theory is to introduce a
massive field $\Sigma$ and a superpotential ${W = \Sigma \left( M
\^ M - \Lambda^4 \right)}$ into the effective action, which thus
takes the form \eqn\SEFF{ S \,=\, \int \! d^4 x \, d^4 \theta \;
K\!\left(M,\bar M;\Sigma,\bar\Sigma\right) \,+\, \int \! d^4 x \,
d^2 \theta \; W \,+\, c.c.,} where $K$ is the K\"ahler potential.
Solving the equations for a critical point of $W$, we find that
$\Sigma=0$ and $M\wedge M=\Lambda^4$. In this description, the
term in the effective action that signals the deformation is
clearly an $F$-term, the term $\Delta W=-\Lambda^4\Sigma$ in the
effective superpotential. If this term is dropped, the constraint
reduces to the classical one $M\wedge M=0$.

The  description we have just given is useful for this particular
example, but relies on being able to describe the  moduli space
${\cal M}_{cl}$ and its deformation ${\cal M}$ in terms of
unconstrained linear fields $\Sigma$ and $m^I$, $I=1,\dots,6$,
together with a superpotential. There are seven of these fields in
all, of which in any given vacuum (away from the origin) two,
namely $\Sigma$ and a linear combination of the $m^I$, are
massive, while five components of $m^I$ are massless and
parametrize the moduli space. Obviously, our deformation $\Delta
W=-\Lambda^4\Sigma$ depends on the massive fields.  In an
analogous but different example, we might be unable to usefully
describe the moduli space in terms of a linear sigma model with a
superpotential. How can we describe the deformation in a low
energy effective action constructed only from the massless fields?

We can find the answer by integrating out the massive fields to
convert $\Delta W$ into an effective interaction for massless
fields only.  In doing so, we work modulo $D$-terms and
attempt to determine what $F$-terms are generated.  This computation
is both simple and instructive and we will perform it, along with an
analogous computation in the theory with six doublets, in section
4.2.

However, for now it is useful to simply use supersymmetry to
determine what $F$-terms are possible on ${\cal M}_{cl}$. At least
away from the origin of ${\cal M}_{cl}$, this theory is
intrinsically described as an $\CN=1$ supersymmetric, nonlinear
sigma model governing maps ${\phi: M^4 \longrightarrow {\cal
M}_{cl}}$ from Minkowski space $M^4$ to ${\cal M}_{cl}$.

From this perspective, the perturbative effective action is the
usual sigma model action, \eqn\SEFFII{ S \,=\, \int \! d^4 x \,
d^4 \theta \; K\!\left(\Phi^i,\bar \Phi{}^{\bar i}\right)\,.} Here
$\Phi^i$ and $\bar\Phi{}^{\bar i}$ are chiral and anti-chiral
superfields whose lowest components $\phi^i$ and $\bar\phi{}^{\bar i}$
are local holomorphic and anti-holomorphic coordinates on ${\cal
M}_{cl}$,\foot{In this discussion, `$i$' is not a flavor index but
an index parametrizing local coordinates on ${\cal M}_{cl}$.} and
$K$ is again the K\"ahler potential associated to some K\"ahler
metric ${ds^2 \,=\, g_{i \bar i} \, d\phi^i d\bar\phi{}^{\bar i}}$ on
${\cal M}_{cl}$.  The reason that we consider a sigma model whose
target is ${\cal M}_{cl}$ is that this is the low energy structure
in perturbation theory.  We want to know how this description may
be modified by instantons, in other words, what $F$-term on ${\cal
M}_{cl}$ may be induced by instantons.

Of course, we know something about the answer: this $F$-term must
describe the deformation of ${\cal M}_{cl}$ into ${\cal M}$. So
let us discuss what terms in the effective action of an ${\cal
N}=1$ sigma model with a given target (in our case, ${\cal
M}_{cl}$) describe a deformation of the complex structure of the
target. We want to consider the deformation not extrinsically, as
a modification of some algebraic equations describing the target,
but intrinsically, as a modification of the $\bar\partial$
operator of the target.

In general, a deformation of the complex structure on ${\cal
M}_{cl}$ is described as a change in the $\bar\partial$ operator
on ${\cal M}_{cl}$ of the form \eqn\DBAR{ \bar\partial_{\bar j}
\,\longmapsto\, \bar\partial_{\bar j} \,+\, \omega_{\bar j}{}^i \,
\partial_i\,.} Here $\omega_{\bar j}{}^i$ is a representative of a
Dolbeault cohomology class in $H^1({\cal M}_{cl},T{\cal M}_{cl})$,
whose elements parametrize infinitesimal deformations of ${\cal
M}_{cl}$.  We use standard notation, with $T{\cal M}_{cl}$ and
$\Omega^1_{{\cal M}_{cl}}$ denoting the holomorphic tangent and
cotangent bundles of ${\cal M}_{cl}$.

We can equally well represent the change \DBAR\ in the
$\bar\partial$ operator on ${\cal M}_{cl}$ as a change in the dual
basis of holomorphic one-forms $d\phi^i$,
\eqn\DPHI{ d\phi^i \,\longmapsto\, d\phi^i \,-\, \omega_{\bar
j}{}^i \, d \bar\phi{}^{\bar j}\,.} As a result, under the deformation,
the metric on ${\cal M}_{cl}$ changes as \eqn\DG{ g_{i \bar i} \,
d\phi^i d\bar\phi{}^{\bar i} \,\longmapsto\, g_{i \bar i} \, \left(
d\phi^i \,-\, \omega_{\bar j}{}^i \, d\bar\phi{}^{\bar j}\right)
d\bar\phi{}^{\bar i}\,,} so that, upon deforming ${\cal M}_{cl}$, the
metric picks up a component of type $(0,2)$ when written in the
original holomorphic and anti-holomorphic coordinates.  (Of
course, there is also a complex conjugate term of type (2,0).)

Since we know how the metric on ${\cal M}_{cl}$ changes when
${\cal M}_{cl}$ is deformed, we can immediately deduce that the
corresponding correction to the sigma model action is generally of
the form \eqn\DSEFF{ \delta S \,=\, \int \! d^4 x \, d^2 \theta \;
\omega_{\bar i\, \bar j} \> \bar D \mskip 2 mu \bar\Phi{}^{\bar i}
\cdot \bar D \mskip 2 mu \bar\Phi{}^{\bar j}\,=\, \int \! d^4 x \;
\omega_{\bar i \, \bar j} \> d\bar\phi{}^{\bar i} \,
d\bar\phi{}^{\bar j} + \cdots\,,} with \eqn\DOM{ \omega_{\bar i
\,\bar j} \,=\, \ha \left(g_{i \bar i} \, \omega_{\bar j}{}^i
\,+\, g_{i \bar j} \, \omega_{\bar i}{}^i\right)\,.} Here $\bar D
\equiv \bar D_{\dot \alpha}$ is the usual spinor covariant
derivative on superspace, and we have introduced the shorthand
notation ``$\cdot$'' for the contraction of spinor indices (so for
any two spinors $\eta$ and $\zeta,$ $\eta\cdot \zeta$ is shorthand
for $\eta_{\dot\alpha}\zeta^{\dot\alpha}$). We have also performed
the fermionic integral over $\theta$ in \DSEFF, from which we see
that the leading bosonic term reproduces the correction to the
metric in \DG.

Of course, the most important property of $\delta S$ --- and the
primary motivation for this paper --- is the fact that $\delta S$
is an $F$-term.   But $\delta S$ is not a correction to the
superpotential -- it generates terms with two derivatives of
bosons, or with four fermions.  Because of the latter
contribution, $\delta S$ is a special case of what we call a
multi-fermion $F$-term.

In contrast to a superpotential interaction, a deformation of the
complex structure (of a smooth complex manifold, such as ${\cal
M}_{cl}$ with the origin removed) is trivial locally.  So locally
on ${\cal M}_{cl}$, it must be possible to write $\delta S$ in the
form $\int d^4\theta(\dots)$.  As will become clear, this cannot
be done globally on ${\cal M}_{cl}$, and it cannot be done even
locally in a way that respects the $SU(4)$ flavor symmetry.  In
that sense, $\delta S$ is a non-trivial $F$-term.

We also note that this $F$-term is not manifestly supersymmetric,
since the operator ${\CO_\omega \,=\, \omega_{\bar i \, \bar j} \>
\bar D \mskip 2 mu \bar\Phi{}^{\bar i} \cdot \bar D \mskip 2 mu
\bar\Phi{}^{\bar j}}$ is not manifestly chiral. Rather, the chirality
of $\CO_\omega$ in the on-shell supersymmetry algebra determined by
the unperturbed sigma model action $S$ follows from the fact that
$\omega_{\bar j}{}^i$ is annihilated by $\bar\partial$.

In section 2.2, we discuss more systematically the basic
properties of multi-fermion $F$-terms such as $\delta S$.

\medskip\noindent{\it Computing $\delta S$ in SQCD}\smallskip

We have described in general what sort of term in the low energy
effective action of an ${\cal N}=1$ sigma model  describes the
deformation of the complex structure of the target.   We will now
be more explicit for $SU(2)$ gauge theory with four doublets.

For this purpose, we reconsider the extrinsic, algebraic
description of the deformation of ${\cal M}_{cl}$, using the
coordinates $m^I$.  Instead of saying that the deformation changes
the equation \eqn\gommo{\sum_I(m^I)^2=0} to
\eqn\tommo{\sum_I(m^I)^2=\epsilon,} we want to work in a
description in which the target space remains the same (away from
the origin and perturbatively in $\epsilon$) but a new interaction
is generated.

It is possible to do this because away from the origin the quantum
constraint \tommo\ can be converted to the classical constraint
\gommo\ by a non-holomorphic change of variables.  When the
coordinates $m^I$ satisfy the quantum constraint \tommo, the new
coordinates  \eqn\NEWM{ \wt m^I \,=\, m^I \,-\, {\epsilon\over 2}
\, {{\delta^I_{\bar J} \, \bar m^{\bar J}} \over {\bar m m}}} obey
the classical constraint \gommo\ to first order in $\epsilon$. (We
could work beyond first order, but this is not necessary.) Here
$\bar m m=\sum_{I=1}^6|m^I|^2$, and in describing $\wt m^I$ we
introduce the tensor $\delta^I_{\bar J}$ constructed from the
$SO(6)$ invariant tensors $\delta^{I J}$ and $\delta_{I \bar J}$;
in the language of $SU(4)$, these tensors would be respectively
$\epsilon^{i j}{}_{\bar k \, \bar l}$, $\epsilon^{i j k l}$, and
$(\delta_{i \bar i} \, \delta_{j \bar j} \,-\, \delta_{j \bar i}
\, \delta_{i \bar j})$.

Thus, when the original coordinates $m^I$ satisfy the quantum
constraint, the new coordinates $\wt m^I$ satisfy the classical
constraint, at least to leading order in $\epsilon$, \eqn\NEWMPL{
\sum_{I=1}^6 \, \left(\wt m^I\right)^2 \,=\, {\cal O}(\epsilon^2)\,.}

The new coordinates $\wt m^I$ are obviously not holomorphic in the
old complex structure on ${\cal M}_{cl}$, but we can find a new
complex structure in which they are holomorphic.  The deformation
of the complex structure can be described as in \DBAR\ by
correcting the $\bar\partial$ operator. From the requirement that
the new coordinates $\wt m^I$ be holomorphic in the new complex
structure, we have that \eqn\HOLNEWM{ \left({\partial \over
{\partial \bar m^{\bar J}}} \,+\, \omega_{\bar J}{}^I \, {\partial
\over {\partial m^I}}\right) \wt m^K \,=\,0\,.}  From this
equation, we can directly solve for the tensor $\omega_{\bar
J}{}^I$ in terms of the components $m^I$ of $M$. We find, again to
leading order in $\epsilon$, that \eqn\BIGOM{ \omega_{\bar J}{}^I
\,=\, { \epsilon\over 2} \left( {{\delta^I_{\bar J}} \over {\bar m
m}} - {{\bar m^I m_{\bar J}} \over {\left(\bar m m\right)^2}} -
{{m^I \bar m_{\bar J}} \over {\left(\bar m m\right)^2}}\right),}
with indices raised and lowered with $\delta^{I J}$ and $\delta_{I
\bar J}$ as appropriate.

In this expression, only the first two terms in \BIGOM\ arise directly
from solving the equation \HOLNEWM.  In fact, the last term in the
expression for $\omega_{\bar J}{}^I \, d\bar m^{\bar J} \, \partial /
\partial m^I$ vanishes identically when we restrict to ${\cal M}$, as
on ${\cal M}$ we have the relation \eqn\COTX{0= \sum_{I=1}^6 \,
\bar m^{\bar I} d \bar m^{\bar I}={1\over 2}d\left(\sum_I(\bar
m^{\bar I})^2\right).} We have included this trivial term in
$\omega_{\bar J}{}^I$ just so that, upon lowering one index with
the K\"ahler metric, the tensor $\omega_{\bar I\, \bar J}$ is
manifestly symmetric.

Of course, we do not actually know the K\"ahler metric $g$ on ${\cal
M}_{cl}$, as appears implicitly in determining $\delta S$ by
converting the section $\omega$ of $\bar\Omega^1_{\CM_{cl}} \otimes
T\CM_{cl}$ to a section of $\bar\Omega^1_{\CM_{cl}} \otimes
\bar\Omega^1_{\CM_{cl}}$, as in \DSEFF\ and \DOM.  By symmetry, we do
know that this metric must equal the metric on ${\cal M}_{cl}$ induced
from the Euclidean metric times a function of $\bar m m$, and
asymptotically for large $\bar m m$ the metric must reduce to the
classical metric describing canonical kinetic terms for underlying
quarks in the ultraviolet regime of SQCD.

All of our expressions for the multi-fermion $F$-terms depend on
the metric $g$. However, this dependence is irrelevant in the
sense that the fundamental holomorphic object $\omega$ which
represents a class in $H^1({\cal M},T{\cal M})$ and determines the
existence of the multi-fermion $F$ term does not depend on a
choice of K\"ahler metric. Of course, the metric is known
asymptotically, near infinity on ${\cal M}$, where it can be
determined from the underlying classical field theory and
asymptotic freedom.

We will now give a concrete formula for $\delta S$. Because of the
dependence on $g$, we can present this formula in various ways.
The most general approach, which also leads to the simplest
expressions, is simply to leave $g$ implicit, absorbing it into
the index structure of $\omega_{\bar I \, \bar J}$ as we did in
\DOM.  This means that we simply use an unknown K\"ahler metric in
raising and lowering indices.  With this convention understood,
from \DSEFF, \DOM, and \BIGOM, we see that $\delta S$ takes the
form \eqn\DSEFFm{ \delta S\,=\, \int \! d^4 x \, d^2 \theta \;
{\epsilon \over 2 } \> \left( {{\delta^{I J}} \over {\bar m m}}
\,-\, {{\bar m^I \, m^J} \over {(\bar m m)^2}} \,-\, {{m^I \, \bar
m^J} \over {(\bar m m)^2}}\right) \> \bar {D} \mskip 2 mu \bar{
m}_I \cdot \bar{D} \mskip 2 mu \bar{ m}_J\,.} Alternatively, this
expression \DSEFFm\ is what results if we assume that $g$ is the
flat metric, so that we simply raise and lower indices with the
Kronecker delta.

On the other hand, because the mesons $m^I$ and $\bar m^{\bar I}$
most naturally (that is in the classical theory) have dimension
$2$, the metric $g_{I \bar I}$ most naturally has dimension $-2$
(so that $ds^2 = \, g_{I \bar I} dm^I dm^{\bar I}$ has dimension
two). As a result, the dimensional analysis of our expression in
\DSEFFm\ is not transparent.  Asymptotically on ${\cal M}$, the
K\"ahler potential is known to be asymptotic to $K=\sqrt{\bar m
m}$.  With this knowledge, we can make the asymptotic form of the
interaction more precise.  In doing so, it is convenient to also
make dimensional analysis manifest by simply using the Kronecker
delta $\delta_{I\bar I}$ to raise and lower indices on $m$ and
$\bar m$, while writing factors of $\sqrt{\bar mm}$ explicitly. In
this case, all components of $m$ and $\bar m$ with indices up or
down have dimension two. The asymptotic form of the interaction
$\delta S$ then  becomes \eqn\DSEFFmII{ \delta S\,=\, \int \! d^4
x \, d^2 \theta \; {\epsilon \over {2 \sqrt{\bar m m}}} \> \left(
{{\delta^{I J}} \over {\bar m m}} \,-\, {{\bar m^I \, m^J} \over
{(\bar m m)^2}} \,-\, {{m^I \, \bar m^J} \over {(\bar m
m)^2}}\right) \> \bar {D} \mskip 2 mu \bar{ m}_I \cdot \bar{D}
\mskip 2 mu \bar{ m}_J\,.} Recalling that $\epsilon \sim
\Lambda^4$, one can check directly that the naive dimensional
analysis holds.

In the rest of the paper, we will mainly follow the first
convention, as in \DSEFFm, so that $g$ appears only implicitly.

In terms of the components $M^{i j}$ of $M$ written using $SU(4)$
flavor indices, as we will use in section 3, the expression
\DSEFFm\ becomes \eqn\DSEFFIII{\eqalign{ \delta S \,=\, \int \!
d^4 x \, d^2 \theta \; &\Lambda^4 \> \left( {{\epsilon^{i_1 j_1
i_2 j_2}} \over {\bar M M}} \,-\, {{\epsilon^{i_1 j_1 k l} \,
M^{i_2 j_2} \, \bar M_{k l}} \over {2 \left(\bar M M\right)^2}}
\,-\, {{\epsilon^{i_2 j_2 k l} \, M^{i_1 j_1} \, \bar M_{k l}}
\over {2 \left(\bar M M\right)^2}} \right) \> \times\cr &\times \>
\bar {D}\mskip 2 mu \bar{ M}_{i_1 j_1} \cdot \bar {D}\mskip 2 mu
\bar{ M}_{i_2 j_2}\,.\cr}}  Here we take $\bar M M \equiv \ha
\lower.8ex\hbox{$\sum\atop{ij}$} \; \bar M_{i j} M^{i j}$.  (The
factor of $1/2$ is included so that if the only nonzero components
of $M^{ij}$ are $M^{12}=-M^{21}=1$, then $\bar MM=1$.  The factors
of $1/2$ in \DSEFFIII\ relative to \DSEFFm\ arise from this
convention and lead to the simple formula below.)

For future reference, we observe that up to a constant factor the
expression in \DSEFFIII\ can be written more compactly as
\eqn\DSEFFIV{\eqalign{ \delta S \,&=\, \Lambda^4\int \! d^4 x \,
d^2 \theta \; \left(\bar M M\right)^{-2} \, \epsilon^{i_1 j_1 i_2
j_2} \, \bar M_{i_1 j_1} \, \left( M^{k l} \, \bar {D}\mskip 2
mu\bar{ M}_{i_2 k} \cdot \bar {D}\mskip 2 mu \bar{M}_{l
j_2}\right)\,.\cr}}
In section 3, we will show that this form
\DSEFFIV\ of the $F$-term is completely determined by symmetry and
furthermore extends naturally to the case of $SU(2)$ SQCD with
$n > 2$ flavors.

\subsec{Multi-Fermion $F$-terms}

Our description of the complex structure deformation in SQCD by
means of a multi-fermion $F$-term may seem perverse, as the
algebraic description of the deformation in \QPLUCKER\ is so much
simpler than \DSEFFIV.  However, by phrasing this deformation as a
multi-fermion $F$-term in an effective four-dimensional $\CN=1$
supersymmetric sigma model, we can see an immediate generalization
to $F$-terms of even higher order.

To introduce this generalization, we begin by recalling that a
four-dimensional sigma model with $\CN=1$ supersymmetry can be
dimensionally reduced to a two-dimensional sigma model with
$\CN=(2,2)$ supersymmetry. Under this reduction, chiral operators
in one sigma model map naturally to chiral operators in the other.
The multi-fermion $F$-terms in four dimensions  have better-known
analogs in two dimensions.

In two dimensions, rings of chiral operators have been much
studied \refs{\LercheUY\Warner\WittenZZ{--}\BershadskyCX} in the
context of string theory and correspond to the rings of local
observables in the topological $A$- and $B$-models.  In fact --
with the superpotential being a typical example -- $F$-terms in
four dimensions reduce to chiral observables of the $B$-model in
two dimensions.  The chiral operators of the $B$-model are in
one-to-one correspondence with elements of $H^p({\cal
M},\bigwedge^qT{\cal M}).$

To construct multi-fermion $F$-terms, we begin with a section
$\omega$ of $\bar\Omega^p_{\cal M} \otimes \bar\Omega^p_{\cal M}$.
(Were it not for the requirement of Lorentz-invariance, we could
more generally start with a section of  $\bar\Omega^p_{\cal M}
\otimes \bar\Omega^q_{\cal M}$ for $p\not= q$.) Explicitly,
$\omega$ is given by a  tensor $\omega_{\bar i_1 \cdots \bar i_p
\,\bar j_1 \cdots \bar j_p}$ that is antisymmetric in the $\bar
i_k$ and also in the $\bar j_k$.  Given such a tensor, we
construct a  possible term in the effective action that
generalizes what we found in  \DSEFF: \eqn\FTERM{\eqalign{ \delta
S \,&=\, \int \! d^4 x \, d^2 \theta \; \omega_{\bar i_1 \cdots
\bar i_p \,\bar j_1 \cdots \bar j_p} \; \left(\bar D \mskip 2
mu\bar\Phi{}^{\bar i_1} \cdot \bar D\mskip 2 mu\bar\Phi{}^{\bar
j_1}\right) \cdots \left(\bar D\mskip 2 mu\bar\Phi{}^{\bar i_p} \cdot
\bar D\mskip 2 mu\bar\Phi^{\bar j_p}\right)\,,\cr
&\equiv\, \int \! d^4 x \, d^2 \theta \; \CO_\omega\,.\cr}} To
achieve Lorentz invariance, spinor indices are contracted here;
for example, $\left(\bar D\mskip 2 mu\bar\Phi{}^{\bar i_1} \cdot \bar
D\mskip 2 mu\bar\Phi{}^{\bar j_1}\right)$ is an abbreviation for
$\left(\bar D_{\dot\alpha}\bar\Phi{}^{\bar i_1} \, \bar D^{\dot
\alpha}\bar\Phi{}^{\bar j_1}\right)$.  Furthermore, given the form of
this operator, we can assume that $\omega$ is symmetric under the
overall exchange of $i$'s and $j$'s.

\medskip\noindent{\it Supersymmetry of $\CO_\omega$}\smallskip

The interaction $\delta S$ is not manifestly supersymmetric.  For
it to be supersymmetric, $\CO_\omega$ must be chiral, that is,
annihilated by the anti-chiral supersymmetries $\bar Q{}_{\dot
\alpha}$. Even if $\delta S$ is supersymmetric, it may  represent
a trivial $F$-term. Though written in \FTERM\ in the form $\int
d^2\theta(\dots)$, it may be that $\delta S$ can be alternatively
written $\int d^4\theta(\dots)$, in other words as a $D$-term.
This will be so if it is possible to write
$\CO_\omega=\{\bar Q{}_{\dot\alpha},[\bar Q{}^{\dot \alpha},V]\}$
for some $V$.  If so, $\CO_\omega$ is trivially chiral and $\delta S=\int
d^4x\,d^4\theta\, V$.

To describe the chirality condition on $\CO_\omega$, which will be
no surprise from experience with the two-dimensional $B$-model, we
first note that we can use the K\"ahler metric $g_{i \bar i}$ on
${\cal M}$ to raise either set of $\bar i$ or $\bar j$ indices on
$\omega$. The raised indices become holomorphic, so upon raising
the indices,  $\omega$ becomes interpreted as  a section of
$\bar\Omega^p_{\cal M} \otimes \bigwedge^p T{\cal M}$ in two
distinct ways.  By our assumption on the symmetry of $\omega$, we
find the same section of $\bar\Omega^p_{\cal M} \otimes
\bigwedge^p T{\cal M}$ either way.

We now consider the action of the anti-chiral supercharges $\bar
Q{}_{\dot \alpha}$ in the on-shell supersymmetry algebra of the
unperturbed sigma model, so that we consider for simplicity only
the linearized supersymmetry constraint on $\delta S$.  Under the
action of $\bar Q{}_{\dot \alpha}$, the component fields $\phi^i$ and
$\psi^i_\beta$ of $\Phi^i$ and the component fields $\bar\phi{}^{\bar i}$
and $\bar\psi{}^{\bar i}_{\dot \beta}$ of $\bar\Phi{}^{\bar i}$
transform as \eqn\EOM{\matrix{&\eqalign{
&\delta_{\dot \alpha} \phi^i \,=\, 0\,,\cr
&\delta_{\dot \alpha} \psi^i_{\beta} \,=\, i \, \partial_{\dot \alpha
\beta} \phi^i\,,\cr}
&\eqalign{
&\delta_{\dot \alpha} \bar\phi{}^{\bar i} \,=\,
\bar\psi{}^{\bar i}_{\dot \alpha}\,,\cr
&\delta_{\dot \alpha} \bar\psi{}^{\bar i}_{\dot \beta} \,=\, -
\Gamma^{\bar i}_{\bar j\,\bar k} \> \bar\psi{}^{\bar j}_{\dot \alpha}
\, \bar\psi{}^{\bar k}_{\dot \beta}\,.\cr}}}
Here $\Gamma$ is the connection associated to the K\"ahler metric
$g_{i \bar i}$ on $\CM$.  So long as we consider only the action of a
single supercharge, we can without loss set $\Gamma$ to zero by a suitable
coordinate choice on $\CM$.

By using the metric to interpret each set of anti-chiral fermions
$\bar\psi{}^{\bar i}_{\dot \beta}$ for $\dot \beta = 1,2$ as
alternatively anti-holomorphic one-forms $d\bar\phi{}^{\bar i}$ or
holomorphic tangent vectors $\partial / \partial \phi^i$, we see
directly from \EOM\ that the action of each of the two supercharges
$\bar Q{}_{\dot \alpha}$ on $\CO_\omega$ corresponds to the action of
$\bar\partial$ on $\omega$ when $\omega$ is regarded as a section of
$\bar\Omega^p_{\cal M} \otimes \bigwedge^p T{\cal M}$ in either of the
two possible ways. Thus, the chirality constraint on $\CO_\omega$ is
simply the condition that $\omega$ be annihilated by $\bar\partial$.
This result is familiar in the $B$-model.

\medskip\noindent{\it Cohomology of $\CO_\omega$}\smallskip

We must also impose an equivalence relation on the space of
operators $\CO_\omega$,  such that $\CO_\omega$ is considered
trivial if $\delta S$ is equivalent to a $D$-term. The condition
we will get is closely related to the reduction to $\bar\partial$
cohomology in the $B$-model.

As a simple example, any perturbative  correction $\delta K$ to
the K\"ahler form can be trivially rewritten as an $F$-term
correction upon performing half the integral over superspace:
\eqn\DSF{\eqalign{ \int \! d^4 x \, d^4 \theta \; \delta K &= \int
\! d^4 x \, d^2 \theta \; \bar D^2 \! \delta K\,,\cr &= \int \!
d^4 x \, d^2 \theta \; \nabla_{\bar i} \nabla_{\bar j} \delta K \,
\left(\bar D\mskip 2 mu\bar\Phi{}^{\bar i} \cdot \bar D\mskip 2
mu\bar\Phi{}^{\bar j}\right).\cr}}
In the second line, we have introduced the covariant derivative
$\nabla$ associated to the connection $\Gamma$ in \EOM, and we
have explicitly rewritten the chiral integrand in the form of an
operator $\CO_\omega$, with \eqn\OK{ \omega_{\bar i\, \bar j} =
\nabla_{\bar i} \nabla_{\bar j} \delta K\,.}

Even more generally, we must consider possible corrections to the
effective action which involve integrals over three quarters of
superspace\foot{We do not know of any actual examples of operators
of this type that can be written as integrals over three quarters
of superspace but not over all of superspace.} and are of the form
\eqn\FDTERM{\eqalign{ \delta S \,&=\, \int \! d^4 x \, d^2 \theta
\, d\bar\theta{}_{\dot \alpha} \; \xi_{\bar i_2 \cdots \bar i_p \,
\bar j_1 \cdots \bar j_p} \; \bar D{}^{\dot \alpha} \bar\Phi{}^{\bar j_1}
\, \left(\bar D\mskip 2 mu\bar \Phi{}^{\bar i_2} \cdot \bar D\mskip 2
mu\bar \Phi{}^{\bar j_2}\right) \cdots \left(\bar D\mskip 2
mu\bar\Phi{}^{\bar i_p} \cdot \bar D\mskip 2 mu\bar\Phi{}^{\bar
j_p}\right)\,,\cr &\equiv\, \int \! d^4 x \, d^2 \theta \,
d\bar\theta{}_{\dot \alpha} \; \CO_\xi{}^{\dot \alpha}\,,\cr &=\,
\int \! d^4 x \, d^2 \theta \; \nabla_{\bar i_1} \xi_{\bar i_2
\cdots \bar i_p\,\bar j_1 \cdots \bar j_p} \; \left(\bar D\mskip 2 mu\bar
\Phi{}^{\bar i_1} \cdot \bar D\mskip 2 mu\bar\Phi{}^{\bar j_1}\right) \cdots
\left(\bar D\mskip 2 mu\bar\Phi{}^{\bar i_p} \cdot \bar D\mskip 2
mu\bar\Phi{}^{\bar j_p}\right)\,.\cr}} Here $\xi$ is a section of
$\bar\Omega^{p-1}_{\cal M} \otimes \bar\Omega^p_{\cal M}$.

Because the correction in \FDTERM\ has the same form as the
$F$-term in \FTERM, we must consider the chiral operators
$\CO_\omega$ as defined up to the equivalence \eqn\EQV{ \CO_\omega
\,\sim\, \CO_\omega \,+\, \left\{\bar Q{}_{\dot \alpha},
\CO_\xi{}^{\dot \alpha}\right\}\,.} Mathematically, this
equivalence becomes an equivalence relation on sections of
$\bar\Omega^p_{\cal M} \otimes \bar\Omega^p_{\cal M}$, \eqn\EQVII{
\omega_{\bar i_1 \cdots \bar i_p \, \bar j_1 \cdots \bar j_p}
\,\sim\, \omega_{\bar i_1 \cdots \bar i_p \, \bar j_1 \cdots \bar
j_p} \,+\, \nabla_{[{\bar i_1}} \xi_{\bar i_2 \cdots {\bar i_p}]
\, \bar j_1 \cdots \bar j_p} \,+\, \left(\bar i_k \leftrightarrow
\bar j_k\right)\,.} As we indicate, the term involving $\xi$ is to
be symmetrized like $\omega$ under the exchange of all pairs $\bar
i_k \leftrightarrow \bar j_k$.

Because of this symmetrization, the equivalence relation implied
by \EQVII\ on sections of $\bar\Omega^p_{\cal M} \otimes
\bigwedge^p T{\cal M}$ is not the same as the usual equivalence
relation in Dolbeault cohomology. Furthermore, since the
corrections \FDTERM\ arise from an integral only over three
quarters of superspace, they are not supersymmetric unless we
impose the (nontrivial) condition that $\bar Q{}^2$ annihilate the
operator $\CO_\xi{}^{\dot \alpha}$, which implies a corresponding
constraint on the sections $\xi$ which appear in \FDTERM\ and
\EQVII.

We are unaware of a more standard mathematical description of this
sort of cohomology, specific to the bundles $\bar\Omega^p_{\cal M}
\otimes \bigwedge^p T{\cal M}$ on an arbitrary K\"ahler manifold,
and we will not comment further on its general structure. Luckily,
symmetries alone will suffice in section 3 to show that the
operators $\CO_\omega$ which we consider for SQCD cannot be
written as integrals over three-fourths of superspace, much less
all of it.

\subsec{Adding a Superpotential to the Sigma Model}

Although we are most interested in SQCD with massless flavors, a
useful technique to study this theory is to consider instead SQCD
with massive flavors and to ask how various observables depend
upon the mass parameters.  Because these mass parameters appear in
a superpotential, holomorphy serves as a powerful tool to
constrain their appearance in the effective action.  In section 4,
we will apply exactly this technique as one way to compute the
multi-fermion $F$-terms in SQCD.

More generally, we can consider adding any background
superpotential $W$ to the basic sigma model action, \eqn\SMW{ S \,
= \, \int \! d^4 x \, d^4 \theta \; K(\Phi^i,\bar\Phi{}^{\bar i})
\,+\, \int \! d^4 x \, d^2 \theta \; W(\Phi^i) \,+\, c.c.} Because
of the superpotential, the on-shell supersymmetry algebra of the
sigma model is altered, and hence the chirality condition on
$\CO_\omega$ is also altered. This fact is fundamental to our
study of the mass deformation of SQCD in section 4, so we pause to
explain it here in the general setting.

In the new action \SMW, the on-shell variations under $\bar
Q{}_{\dot \alpha}$ of the component fields $\phi^i$,
$\bar\phi{}^{\bar i}$, $\psi^i_\beta$, and $\bar\psi{}^{\bar
i}_{\dot \beta}$ are now given by \eqn\EOMII{\matrix{&\eqalign{
&\delta_{\dot \alpha} \phi^i \,=\, 0\,,\cr &\delta_{\dot \alpha}
\psi^i_{\beta} \,=\, i \, \partial_{\dot \alpha \beta}
\phi^i\,,\cr} &\eqalign{ &\delta_{\dot \alpha} \bar\phi{}^{\bar i}
\,=\, \bar\psi{}^{\bar i}_{\dot \alpha}\,,\cr &\delta_{\dot
\alpha} \bar\psi{}^{\bar i}_{\dot \beta} \,=\, - \Gamma^{\bar
i}_{\bar j\,\bar k} \> \bar\psi{}^{\bar j}_{\dot \alpha} \,
\bar\psi{}^{\bar k}_{\dot \beta}\,+\, \epsilon_{\dot \alpha \dot
\beta} \, g^{\bar i i} \partial_i W\,.\cr}}} Because of the
appearance of the one-form $dW$ in the variation of
$\bar\psi{}^{\bar i}_{\dot \beta}$ in \EOMII, the action of the
supercharges $\bar Q{}_{\dot \alpha}$ on $\CO_\omega$ is no longer
given geometrically by the action of $\bar\partial$ on $\omega$.
Instead, when $\bar\psi{}^{\bar i}_{\dot \beta}$ is interpreted as
a holomorphic tangent vector $\partial / \partial \phi^i$, the
term involving $W$ corresponds geometrically to the interior
product of $\partial / \partial \phi^i$ with the holomorphic
one-form $dW$.  So the $\bar\partial$ operator is now generalized
to the operator \eqn\BD{ \delta = \bar\partial \,+\,
\iota_{dW}\,,} acting on sections of $\bar\Omega^p_{\cal M}
\otimes \bigwedge^p T{\cal M}$.  Here $\iota_{dW}$ denotes the
operator on $\bar\Omega^p_{\cal M} \otimes \bigwedge^p T{\cal M}$
which acts by the interior product with the one-form $dW$.  (In
other words, $\iota_{dW}$ acts by removing $\bar \psi$ and
replacing it with $dW$.)  We note that because $W$ is holomorphic,
$\delta^2 = 0$. Thus, the first order chirality condition on the
operator $\CO_\omega$ becomes the requirement that $\delta$
annihilate $\omega$.

A nice mathematical discussion of the cohomology theory associated
to $\delta$ is given by Liu in \Liu, and applications to string
theory are discussed in \BeasleyFX.

When $\omega$ is a section of $\bar\Omega^1_{\cal M} \otimes T{\cal
M}$, then the modified chirality condition has a
very direct geometric interpretation.  In this case, the condition
that $\delta \omega = 0$ implies that $\omega$ is annihilated
separately by both the operators $\bar\partial$ and $\iota_{dW}$.  The
latter condition implies that \eqn\CHROVI{ \omega_{\bar j}{}^i \,
\partial_i W \,=\, 0\,.}  Since $W$ is holomorphic, this condition is
then equivalent to the condition that \eqn\HOLW{ \left(
\bar\partial_{\bar j} \,+\, \omega_{\bar j}{}^i \, \partial_i
\right)\!W \,=\,0\,,} implying that the deformation of
$\bar\partial$ represented by $\omega$ must preserve the
holomorphy of $W$.  More generally, if it is possible to modify
$W$ to a function $W+\Delta W$ that is holomorphic in the deformed
complex structure, then $\omega+\Delta W$ is annihilated by
$\delta$.

\newsec{Multi-Fermion $F$-Terms in $SU(2)$ SQCD}

Up to this point, we have discussed general properties of
multi-fermion $F$-terms in an arbitrary $\CN=1$ sigma model.  We
now specialize our analysis to the particular case of SQCD.  Our
main goal in the rest of the paper, concentrating mainly on the
example of gauge group $SU(2)$, is to show that multi-fermion
$F$-terms are generated in the effective action of SQCD.

To this end, we begin in this section by analyzing the constraints
imposed by symmetries and holomorphy on the form of any
multi-fermion $F$-term corrections in $SU(2)$ SQCD.  The case of
SQCD with gauge group $SU(2)$ is particularly simple due to the
enhancement of the flavor symmetry.  In this case, we fix the form
of the operators $\CO_\omega$ uniquely, and we demonstrate that
they are nontrivial in the cohomology of $\bar Q{}_{\dot \alpha}$.

In the general case of SQCD with gauge group $SU(N_c)$ and $N_f >
N_c$ flavors, a similar analysis to determine the form of the
operators $\CO_\omega$ appears to be more complicated, since the
geometry of the moduli space ${\cal M}$ itself is more
complicated.  However, the direct instanton computation of section
4.1 shows that such interactions arise for all $N_c$ and $N_f\geq
N_c-1$.  The other derivations in section 4 generalize in spirit.

In the case of $SU(2)$ SQCD with $N_f = n$ flavors, we have
already described algebraically the classical moduli space ${\cal
M}$ as being parametrized by the mesons $M^{i j}$, subject to the
system of quadratic equations ${M \^ M = 0}$.  This description of
${\cal M}$ has the virtue of being very succinct.  However, we now
give another description of ${\cal M}$ which makes its symmetry
more apparent and consequently enables us to determine immediately
the chiral operators $\CO_\omega$ which arise from cohomology
classes on ${\cal M}$.

\subsec{More About the Geometry of ${\cal M}$}

Since symmetries are of the utmost importance, we first review the
symmetries of $SU(2)$ SQCD with $N_f = n$ flavors.  Besides the
$SU(2)$ color and $SU(2n)$ flavor symmetries, this gauge theory
also possesses a non-anomalous $U(1)$ $\CR$-symmetry as well as an
anomalous $U(1)$ axial symmetry.  Under these symmetries, the
quark superfields $Q^i_a$, the mesons $M^{i j}$, and the
holomorphic coupling scale $\Lambda$ transform as follows:
\eqn\SYMS{\matrix{
&\quad&SU(2)_c\quad&SU(2n)\quad&U(1)_A\quad&U(1)_\CR\quad\cr
\noalign{\vskip 2 pt} &Q^i_a\quad&{\bf 2}\quad&{\bf
2n}\quad&1\quad&{1 - {2 \over n}}\quad\cr \noalign{\vskip 1 pt}
&M^{ij}\quad&{\bf 1}\quad&\bigwedge^2\!\left({\bf
2n}\right)\quad&2\quad&2\left({1 - {2 \over n}}\right)\quad\cr
\noalign{\vskip 1 pt} &\Lambda^{6-n}\quad&{\bf 1}\quad&{\bf
1}\quad&2n\quad&0\,.\cr}} Here $\Lambda^{6-n}$ is the standard
instanton counting parameter.  (In this one place, we denote the
gauge group as $SU(2)_c$, to distinguish it from an unbroken
$SU(2)$ flavor group that will appear momentarily.)

We now describe ${\cal M}$ by considering the pattern of symmetry
breaking around a fixed supersymmetric vacuum.  Up to the action
of the symmetries, any solution of the usual $D$-term equations
takes the form \eqn\SDTERM{ Q^i_a \,=\,
\pmatrix{v&0\cr0&v\cr0&0\cr\vdots&\vdots\cr0&0\cr}\,\equiv\, v \,
\hat\delta^i_a\,,} with $v$ being an arbitrary complex number.

So long as $v$ is non-zero, the expectation value of $Q^i_a$ in
\SDTERM\ breaks the symmetry group in \SYMS\ down to a subgroup
\eqn\SUBGP{ SU(2) \times SU(2n-2) \times U(1)_A' \times U(1)_\CR'\,.}
The unbroken $SU(2) \times SU(2n-2)$ factor arises in the obvious way,
and the unbroken $U(1)$ axial and $\CR$-symmetries arise from linear
combinations of the corresponding generators in \SYMS\ with the
diagonal flavor generator in the center of the subgroup
\eqn\SUBGPII{ S\big(U(2) \times U(2n-2)\big) \, \subset\, SU(2n)\,.}

Of course, the gauge group is completely Higgsed, and the massless
fluctuations of the quarks $Q^i_a$ about the point \SDTERM\
decompose into two irreducible representations of the unbroken
symmetry group \SUBGP, with \eqn\SYMSII{\matrix{
&\quad&SU(2)\quad&SU(2n-2)\quad&U(1)_A'\quad&U(1)_\CR'\quad\cr
\noalign{\vskip 2 pt} &\Phi^s_c\quad&{\bf 2}\quad&{\bf
2n-2}\quad&{n \over {n-1}}\quad&{{n-2} \over {n-1}}\quad\cr
\noalign{\vskip 1 pt} &\Phi\quad&{\bf 1}\quad&{\bf
1}\quad&0\quad&0\quad\cr\noalign{\vskip 1pt}
&\Lambda^{6-n}\quad&{\bf 1}\quad&{\bf 1}\quad&2n\quad&0\,.\cr}}
Here the singlet $\Phi$ describes a rescaling of $M$; the other
fluctuations transform as an irreducible representation $\Phi^s_c$
of the unbroken symmetry, where $c=1,2$ is an index labelling the
${\bf 2}$ of the unbroken $SU(2)$ and $s=3,\ldots,2n$ is now an index
labelling the ${\bf 2n-2}$ of $SU(2n-2)$.  Throughout the paper, we
will apply the convention that $c,d,e,f$ refer to indices
$1,2$ of the unbroken $SU(2)$, that $s,t,u,v$ refer to indices
$3,\ldots,2n$ of the unbroken $SU(2n-2)$, and that $i,j,k,l$ run over
all indices $1,\ldots,2n$ of the full $SU(2n)$ flavor symmetry.  These
massless fluctuations $\Phi$ and $\Phi^s_c$ represent local
coordinates on ${\cal M}$, such as were used in section 2.  Finally,
for future reference in section 3.2 we have included in \SYMSII\ the
charges of $\Lambda^{6-n}$, which are identical to those in \SYMS.

Because any solution of the $D$-term equations can be brought to
the form \SDTERM\ using the $SU(2) \times SU(2n)$ symmetry of
SQCD, we see that the $SU(2n)$ flavor symmetry acts transitively
on the quotient of $\CM$ minus the origin by the $\BC^*$ action which
scales $v$.  We thus set $\wt \CM = \CM - \{ 0 \}$, and we let $B$ be this
quotient of $\wt\CM$ by $\BC^*$.

Furthermore, our description of the symmetry breaking pattern in
\SUBGP\ is equivalent to the geometric observation that, at any
non-zero $v$, the subgroup of $SU(2n)$ which stabilizes the point
corresponding to $Q^i_a = v \, \hat\delta^i_a$ on ${\cal M}$ is
$S\big(U(2) \times U(2n-2)\big)$.  Thus, we can describe $B$ as a
homogeneous (and in fact symmetric) space,
\eqn\COS{ B = SU(2n) / S\big(U(2) \times U(2n-2)\big)\,.}

To incorporate the value of $v$ into our description of ${\cal
M}$, we observe that the $\BC^*$ action which scales $v$ is the
complexification of the $U(1)_A$ symmetry in \SYMS.  This symmetry
corresponds to the action of the central $U(1)$ which lies in the
stabilizer subgroup $S\big(U(2) \times U(2n-2)\big)$ and whose
generator mixes with the generator of $U(1)_A$ under the symmetry
breaking.  Associated to this $U(1)$ generator in $S\big(U(2)
\times U(2n-2)\big)$ is a corresponding homogeneous line bundle
$\CL$ --- and hence a $\BC^*$ bundle --- over $B$.  To specify
$\CL$, we simply note that the singlet field $\Phi$ transforms as
a section of $\CL$ and has charge $+2$ under the original $U(1)_A$
symmetry, as $\Phi$ describes the rescaling of $M^{ij}$.

So, if we excise the singularity at the origin of ${\cal M}$, then
$\wt {\cal M}$ can be globally described as this $\BC^*$ bundle over
the base $B$, \eqn\MODX{ \BC^* \longrightarrow \wt {\cal M}
\buildrel\pi\over\longrightarrow B\,.}

A direct relationship now exists between the algebraic description
of ${\cal M}$ in \PLUCKER\ and the intrinsic description of ${\cal
M}$ in \MODX.  To describe this relation, we consider the mesons
$M^{i j}$ modulo overall scaling, corresponding to the $\BC^*$
action generated by $U(1)_A$.  Then the equations ${M \^ M = 0}$
are the classical Pl\"ucker relations \Griffiths\ which describe
the Grassmannian $Gr(2,2n)$ of complex two planes in $\BC^{2n}$ as
an algebraic subvariety of the projective space parametrized by
$M^{i j}$.

On the other hand, this Grassmannian can also be described as a
quotient,
\eqn\GR{ Gr(2,2n) = U(2n) / \big(U(2) \times U(2n-2)\big)\,,}
which is equivalent to our description in \COS\ of the base $B$.
Thus, the $\BC^*$ bundle over $B$ in \MODX\ is simply the bundle
associated to the affine cone over the Grassmannian $Gr(2,2n)$ with
its Pl\"ucker embedding in projective space.  Equivalently, the line
bundle $\CL$ arises as the pullback from the degree one bundle
$\CO(1)$ on projective space.

\subsec{The New $F$-Terms}

With our thorough discussion of the symmetries of SQCD, we can
immediately derive the form of any multi-fermion $F$-terms that
might appear on ${\cal M}$.  We perform our analysis in two steps:
first locally, and then globally.

\medskip\noindent{\it Local Analysis}\smallskip

Locally, we construct the chiral operator $\CO_\omega$ from the
massless fluctuations described by $\Phi^s_c$ and $\Phi$ about the
vacuum $Q^i_a = \hat\delta^i_a$.  Thus, in terms of the section
$\omega$ of $\bar\Omega^p_{\cal M} \otimes \bigwedge^p T{\cal M}$,
we only consider $\omega$ as restricted to the tangent space of
${\cal M}$ at this point.

Now, the operator $\CO_\omega$ must be invariant under the
symmetries ${SU(2) \times SU(2n-2)} \times U(1)_A'$ and must have
charge $+2$ under $U(1)_\CR'$ in \SYMSII.  Furthermore, since we
are only considering the corresponding section $\omega$ as
restricted to the tangent space of a point in $\CM$, we must
construct $\CO_\omega$ completely from the fermionic fields $\bar
D_{\dot \alpha} \bar \Phi$ and $\bar D_{\dot \alpha} \bar
\Phi{}^c_s$ which represent either one-forms or (by raising an
index) tangent vectors to $\CM$.  From \SYMSII\ we see that $\bar
D_{\dot \alpha} \bar \Phi$ and $\bar D_{\dot \alpha} \bar
\Phi{}^c_s$ have respective charges $+1$ and $+1/(n-1)$ under
$U(1)_\CR'$.  So just to make an operator of $U(1)_\CR'$ charge
$+2$, we require that it contain either two copies of $\bar
D_{\dot \alpha} \bar \Phi$, or one copy of $\bar D_{\dot \alpha}
\bar \Phi$ and $n-1$ copies of $\bar D_{\dot \alpha} \bar
\Phi{}^c_s$, or $2(n-1)$ copies of $\bar D_{\dot \alpha} \bar
\Phi{}^c_s$.

We can immediately rule out the first possibility, necessarily of
the form $\bar D\mskip 2 mu\bar \Phi \cdot \bar D\mskip 2 mu\bar
\Phi$, since from \SYMSII\ this operator is not charged under
$U(1)_A'$ and hence is not multiplied by any power of $\Lambda$,
contradicting the fact that our operator must vanish in the
appropriate weak coupling limit as well as the fact that we expect it
to be generated by instantons.  (A more detailed study shows that there are no
non-trivial chiral operators of this type.)  On the other hand,
since the only tensors of $SU(2) \times SU(2n-2)$ which we can use
to make invariants out of the fields $\bar D_{\dot \alpha} \bar
\Phi{}^c_s$ are the anti-symmetric tensors $\epsilon_{c d}$ and
$\epsilon^{s_1 t_1 \cdots s_p t_p}$ with $p=n-1$, we cannot make
an invariant operator from one copy of $\bar D_{\dot \alpha} \bar
\Phi$ and only $n-1$ copies of $\bar D_{\dot \alpha} \bar
\Phi{}^c_s$.

We are left to consider the operator $\CO_\omega$ which is made
from $2(n-1)$ copies of $\bar D_{\dot \alpha} \,\bar\Phi{}^c_s$, of the form
\eqn\LCOM{ \Lambda^{6-n} \> \epsilon^{s_1 t_1 \cdots s_p t_p} \,
\epsilon_{c_1 d_1} \cdots \epsilon_{c_p d_p} \, \left(\bar {D
}\mskip 2 mu\bar\Phi{}_{s_1}^{c_1} \cdot \bar {D}\mskip 2 mu\bar
\Phi{}_{t_1}^{d_1}\right) \cdots \left(\bar {D}\mskip 2 mu\bar
\Phi{}_{s_p}^{c_p} \cdot \bar {D}\mskip 2
mu\bar\Phi{}_{t_p}^{d_p}\right)\,,\quad p = n - 1\,.} This operator is
invariant under $SU(2) \times SU(2n-2)$ and carries charge $+2$
under $U(1)_\CR'$.  The pattern of contractions of spinor indices
is fixed by the fact that each expression in parentheses must be
antisymmetric under exchanges of both the pairs $(c,d)$ and
$(s,t)$ and must also obey Fermi statistics.

Also, we see from \SYMSII\ that each fermion appearing in $\CO_\omega$
carries charge $-n/(n-1)$ under $U(1)_A'$, so the fermionic part of
$\CO_\omega$ carries axial charge $-2n$.  The fact that $\CO_\omega$
must be invariant under the axial symmetry then fixes the dependence
on $\Lambda$.  In particular, we see that the operator in \LCOM\
involves a single power of the instanton counting parameter
$\Lambda^{6-n}$ and so could arise as a one-instanton effect.

So the local form of $\CO_\omega$ is fixed completely by the
symmetries, and moreover $\CO_\omega$ has the correct dependence
on $\Lambda$ to be generated by instantons.  Furthermore, in terms
of the section $\omega$ of $\bar\Omega^p_{\cal M} \otimes
\bigwedge^p T{\cal M}$, we see that the parameter $p$ is related
to the number of flavors $n$ by ${p = n - 1}$.  This fact is a
special case of the relation ${p = N_f - N_c + 1}$ which must hold
in $SU(N_c)$ SQCD with $N_f$ flavors.    In  the direct instanton
computation in section 4,  this relation follows most immediately
by counting fermion zero modes in the instanton background.

\medskip\noindent{\it A Geometric Remark on Pullbacks From
$B$}\smallskip

Because $\CO_\omega$ only involves $\bar {D}\,\bar \Phi{}^c_s$ and
not the singlet $\bar {D}\, \bar\Phi$, the section $\omega$ has
only components along the base $B$, with no legs along the $\BC^*$
fiber. Naively, one might have concluded that $\omega$ then arises
as the pullback from a section of ${\bar\Omega^{n-1}_B \otimes
\bigwedge^{n-1} TB}$ on $B$. Actually, the dependence of
$\CO_\omega$ on scaling of the quark superfields  means that it is
a pullback from a section of ${\bar\Omega^{n-1}_B \otimes
\bigwedge^{n-1} TB}\otimes {\cal L}^k$ for some $k$.\foot{There is
a further inessential subtlety.  A section of $TB$ cannot quite be
pulled back to a section of $T{\cal M}$ as there is a nontrivial
exact sequence $0\to TF\to T{\cal M}\to TB\to 0$ where $TF$ is the
tangent space to the fibers of ${\cal M}\to B$.  Because the
relevant cohomology of $TF$ is trivial, this distinction is
unimportant.}

In fact, $k=-n$.  Indeed, as we noted above, the fermionic part of
${\cal O}_\omega$ carries $U(1)_A'$ charge $-2n$.  As $U(1)_A'$
differs from $U(1)_A$ by a generator of $SU(2n)$ under which
${\cal O}_\omega$ is invariant, this means that, if we omit the
factor of $\Lambda^{6-n}$ from \LCOM, then ${\cal O}_\omega$ has
$U(1)_A$ charge $-2n$.  Since the basic meson field $M$ has $U(1)_A$
charge $2$, this means that $\CO_\omega$ transforms as
$M^{-n}$ and $\omega$ can be regarded as a section of
$\bar\Omega{}^{n-1}_B \otimes \wedge^{n-1} TB\otimes {\cal L}^{-n}$.

Consider a general scaling $M\to \lambda M$, $\bar M\to \bar
\lambda \,\bar M$, for $\lambda\in {\Bbb C}^*$.  Under this
scaling, $\omega\to\lambda^{-n}\bar\lambda{}^0\,\omega=\lambda^{-n}\omega$.
The fact that the exponent of $\bar\lambda$ is zero is implied by
the fact that $\bar\partial \omega=0$, and the fact that the
exponent of $\lambda$ is $-n$ is equivalent to the fact that
$\omega$ is a section of $\bar\Omega{}^{n-1}_B\otimes \wedge^{n-1}
TB\otimes {\cal L}^{-n}$.  We apply these observations when we
write a global expression for $\CO_\omega$.

\medskip\noindent{\it Chirality and Cohomology of
$\CO_\omega$}\smallskip

Let us now check that $\CO_\omega$ is chiral -- annihilated by
$\bar Q{}_{\dot \alpha}$ -- and moreover  represents a nontrivial
$\bar Q{}_{\dot \alpha}$ cohomology class. This check follows
directly from symmetries.

We recall that the chirality condition on $\CO_\omega$ is
equivalent to the geometric condition that $\bar\partial$
annihilate $\omega$. Because $\CO_\omega$ is a pullback from $B$,
we can consider just the action of the $\bar\partial$ operator
along $B$ on $\omega$, considered as a section of $\bar\Omega^p_B
\otimes \bigwedge^p TB \otimes \CL^{-n}$.  Because both the
$\bar\partial$ operator on $B$ and $\omega$ are singlets under the
action of $SU(2) \times SU(2n-2)$, the section $\bar\partial
\omega$ of $\Omega^{p+1}_B \otimes \bigwedge^p TB \otimes
\CL^{-n}$ must also be a singlet. But no (nontrivial) invariant
section of $\Omega^{p+1}_B \otimes \bigwedge^p TB \otimes
\CL^{-n}$ exists; such a section  would be constructed from an
$SU(2)$ singlet made from the tensor product of $2p+1$ ${\bf
2}$'s. So the $\bar\partial$ operator on $B$ necessarily
annihilates $\omega$.

A similar argument based upon symmetries also shows that
$\CO_\omega$ cannot be written in the form $\{\bar Q_{\dot
\alpha}, \CO_\xi{}^{\dot \alpha}\}$ in a way that respects the
flavor symmetry. Indeed, invariant sections of $\bar\Omega^{p-1}_B
\otimes \bigwedge^p TB \otimes \CL^{-n}$ and $\bar\Omega^p_B
\otimes \bigwedge^{p-1} TB \otimes \CL^{-n}$ do not exist, since
one cannot make an $SU(2)$ invariant from $2p-1$ ${\bf 2}$'s.

\medskip\noindent{\it Global Analysis}\smallskip

Our expression in \LCOM\ is only a local expression for
$\CO_\omega$, but because the $SU(2n)$ flavor symmetry acts
transitively on ${\cal M}$, this local expression suffices to
determine a global expression for $\CO_\omega$.  In order to write
such an expression using the mesons $M^{i j}$, we observe that the
local tensors $\epsilon^{s_1 t_1 \cdots s_p t_p}$ and $\epsilon_{c
d}$ in \LCOM\ extend globally to tensors on ${\cal M}$ given by
$\epsilon^{i_1 j_1 \cdots i_n j_n} \, \bar M_{i_1 j_1}$ and $M^{k
l}$.  Then $\CO_\omega$ must take the global form \eqn\COMII{
\CO_\omega \,=\, \Lambda^{6-n} \, F(\bar M M) \, \epsilon^{i_1 j_1
\cdots i_n j_n} \, \bar M_{i_1 j_1} \, \CO_{i_2 j_2} \cdots
\CO_{i_n j_n}\,,} with \eqn\COMIII{ \CO_{ij} \,\equiv\, M^{kl} \,
\bar D \,\bar M_{ik} \cdot \bar D \,\bar M_{lj}\,,\qquad \bar M M
\,\equiv\, \ha \, \bar M_{ij} M^{ij}\,.}  Of course, we
employ the usual summation convention in writing $\bar M M$ as in
\COMIII, using the K\"ahler metric $g$ on ${\cal M}$ to raise and
lower indices throughout.

In writing $\CO_\omega$, we have also included as a prefactor an
invariant function $F(\bar M M)$ on ${\cal M}$ which is not directly
determined by the local expression in \LCOM.  The function $F(\bar M
M)$ is, however, determined by dimensional analysis and also, as we
will now discuss, by requiring ${\cal O}_\omega$ to be chiral.

The chirality condition on $\CO_\omega$ is most naturally expressed as
the condition that the corresponding section $\omega$ of
$\bar\Omega^{n-1}_{\cal M} \otimes \bigwedge^{n-1} T{\cal M}$ be
annihilated by $\bar\partial$.  Explicitly, the section $\omega$ which
determines the operator $\CO_\omega$ in \COMII\ is given globally by
\eqn\SECT{ \omega \,=\, F(\bar M M) \, \epsilon^{i_1 j_1 \cdots
i_n j_n} \bar M_{i_1 j_1} \, \left( M^{k_2 l_2} \, d \bar M_{i_2
k_2} \, {\partial \over {\partial M^{l_2 j_2}}} \right) \cdots
\left( M^{k_n l_n} \, d \bar M_{i_n k_n} \, {\partial \over
{\partial M^{l_n j_n}}} \right)\,.}
In order that $\omega$ be annihilated by $\bar\partial$, we have
already observed that it must be invariant under the scaling $\bar
M\to \bar \lambda \,\bar M$.  Furthermore, in order that $\omega$
arise from a section of the bundle $\bar\Omega^{n-1}\otimes
\wedge^{n-1} TB\otimes {\cal L}^{-n}$, we have also observed that it
must transform under the scaling $M\to\lambda M$ as $\omega \to
\lambda^{-n}\omega$.

However, if we ignore $F(\bar M M)$, we see that $\omega$ in
\SECT\ otherwise scales with degree $n$ in $\bar\lambda$ and with
degree zero in $\lambda$.  Thus, we set $F(\bar M M)=(\bar M
M)^{-n}$ to ensure that $\omega$ scales as $M^{-n}$.  So we must
set \eqn\COMIV{ \CO_\omega \,=\, \Lambda^{6-n} \, \left(\bar M
M\right)^{-n} \, \epsilon^{i_1 j_1 \cdots i_n j_n} \, \bar M_{i_1
j_1} \, \CO_{i_2 j_2} \cdots \CO_{i_n j_n}\,.} This expression
directly generalizes our previous formula \DSEFFIV\ in the special
case $n = 2$.

Let us also make a remark about the global form of $\CO_\omega$, or
equivalently $\omega$ in \SECT.  In this expression, the components
$M^{i j}$ of $M$ which appear are just affine coordinates on a vector
space in which $\CM$ is embedded, and it must be that only the components of
$\partial / \partial M^{i j}$ and $d\bar M{}^{i j}$ which represent tangent
and cotangent vectors to $\CM$ itself appear in \SECT.  To check this
condition, we can without loss consider the point of $\CM$ at which
$M^{i j} = \hat\epsilon^{i j}$.  (We recall that the nonzero
components of $\hat \epsilon$ are
$\hat\epsilon^{12}=-\hat\epsilon^{21}=1$.)  Then the holomorphic
tangent space to $\CM$ at this point is spanned by vectors
$\partial / \partial M^{i j}$ for which both $i,j = 1,2$, corresponding to the
singlet $\Phi$, or for which $i=1,2$ and $j > 2$, corresponding to
$\Phi^s_c$ in the representation $({\bf 2},{\bf 2n -2})$.

In particular, the vector $\partial / \partial M^{i j}$ for which
both $i,j >2$ is not a tangent vector to $\CM$ at this point. So
in order for \SECT\ to be well defined as a section of
${\bar\Omega^{n-1}_{\cal M} \otimes \bigwedge^{n-1} T{\cal M}}$,
such components of $d\bar M_{i j}$ and $\partial / \partial M^{i
j}$ with both $i,j >2$ must not appear.  However, upon
substituting $M^{i j} = \hat\epsilon^{i j}$ into \SECT, we see
that the factors of $\bar M_{i_1 j_1}$ and $M^{k l}$ ensure that
these unwanted components do not appear, and the expression in
\SECT\ is a section of ${\bar\Omega^{n-1}_{\cal M} \otimes
\bigwedge^{n-1} T{\cal M}}$ as claimed.

Like \DSEFFm, \COMIV\ is written in terms of an arbitrary unknown
K\"ahler metric on ${\cal M}$.  As in \DSEFFmII, we can make the
asymptotic behavior more explicit, since we know the asymptotic
form of the K\"ahler metric.  In writing this formula, just as in
\DSEFFmII, we use Kronecker deltas to raise and lower indices on
$M$ (so all components of $M$ and $\bar M$ with index up or down
have dimension two), and write all factors of $\bar M M$
explicitly. With this understood, the asymptotic form of the
interaction is \eqn\jurgo{ \Lambda^{6-n} \, \left(\bar M
M\right)^{-(3n-1)/2} \, \epsilon^{i_1 j_1 \cdots i_n j_n} \, \bar
M_{i_1 j_1} \, \CO_{i_2 j_2} \cdots \CO_{i_n j_n}.}

\newsec{Computing The Multi-Fermion $F$-Terms}

Although symmetries suffice to fix the form of the $F$-term
correction in SQCD uniquely, we must still check that it is
actually generated.  So in this section, we provide three
computations which show this.

\subsec{A Direct Instanton Computation}

Since instanton effects are the subject of the paper, we first
generate the $F$-terms directly by a one-instanton computation
which generalizes the classic one-instanton computation
\refs{\AffleckMK,\CordesUM,\FinnellDR} of the superpotential in
the theory with $N_f = N_c - 1$ flavors.

The most basic, and most illuminating, feature of this instanton
computation is that it directly explains how the relation $p = n - 1$
arises in the $SU(2)$ theory with $N_f = n$ flavors.  This relation
arises from counting fermion zero modes in the instanton background,
and the same counting implies that, in the $SU(N_c)$ theory, we must
have $p = N_f - N_c + 1$.

Very briefly, before we review the details of the instanton
computation, we will explain the counting of fermion zero modes
that controls the structure of the $F$-term.  We thus recall that,
in the one-instanton background, we find at leading order $2 N_c$
gaugino zero modes and $2 N_f$ quark zero modes.  However, beyond
leading order, the Yukawa couplings pair $2(N_c - 1)$ of the
gaugino and quark zero modes, and these modes are lifted.  As a
result, two gaugino zero modes and $2(N_f - N_c + 1)$ quark zero
modes remain. The two gaugino zero modes that remain are generated
by exact global supersymmetries.  Thus, if we consider the general
form of the multi-fermion $F$-term in \FTERM, the two gaugino zero
modes are associated to the fermionic collective coordinates
$\theta^\alpha$ that appear in the integral over superspace, and
the $2(N_f - N_c + 1)$ quark zero modes must be absorbed by the
chiral operator $\CO_\omega$ itself.  So $p = N_f - N_c + 1$.

We now present the details of the instanton computation in the case of
$SU(2)$ SQCD.  As described above, this computation should generalize
directly to the case of $SU(N_c)$ SQCD, though one must consider a
more involved integral over the collective coordinates of the
instanton.

Following closely the computation of Affleck, Dine, and Seiberg
\AffleckMK, we work on the Higgs branch of SQCD, under the
assumption that the classical quark vacuum expectation value,
$Q^i_a = v \, \hat\delta^i_a$, is large and the effective gauge
coupling $g^2(v)$ is small.  In this regime, the approximate
instanton equations are valid, \eqn\INST{ D^\mu F_{\mu \nu} \,=\,
0\,,\quad D^2 q^i_a \,=\, 0\,,} where we recall that $q^i_a$ is
the scalar component of $Q^i_a$.  In a one-instanton background,
the solution of \INST\ for $q^i_a$ with boundary condition fixed
by its classical expectation value is given by \eqn\INSTB{ q^i_a =
{{\sigma_{\mu a}^i\,x^\mu\,v } \over {\sqrt{\rho^2 + x^2}}}\,.}
Here $\sigma_\mu = (1,-i \sigma^A)$, with $\sigma^A$ the Pauli
matrices, are the usual quaternion representatives.  Also, $x^\mu$
is a coordinate on $\BR^4$, and $\rho$ is the scale of the
instanton solution.  The classical action for this instanton
background is \eqn\ACTI{ S_0 = {1 \over {g^2}} \left( 8 \pi^2 + 4
\pi^2 \rho^2 |v|^2\right)\,.} When $|v|^2 \neq 0$, instantons of
large size are exponentially suppressed by this classical action,
and the integral over the scale $\rho$ will be convergent.

We must now consider what sort of correlation function to compute
in order to probe for the multi-fermion $F$-term determined by the
operator $\CO_\omega$ in \COMIV.  For this purpose, we recall the
chiral superfields $\Phi$ and $\Phi^s_c$ which we introduced in
section 3 to describe massless fluctuations of the quark
superfields around the Higgs vacuum. Introducing components for
these fields, \eqn\COMPPHI{\eqalign{ \Phi \,&=\, \phi \,+\, \theta
\chi \,+\, \ldots\,,\cr \Phi^s_c \,&=\, \phi^s_c \,+\, \theta
\chi^s_c \,+\, \ldots\,,\cr}} we see that among the various
interactions which arise from the multi-fermion $F$-term is an
effective interaction for $2n$ fermions of the form \eqn\VRTX{
{{\Lambda^{6-n}} \over {v^4 \, |v|^{2(n-1)}}} \, \int \! d^4 x\;
\epsilon^{s_1 t_1 \cdots s_p t_p} \, \epsilon_{c_1 d_1} \cdots
\epsilon_{c_p d_p} \, \chi \cdot \chi \, \left(\bar
\chi{}_{s_1}^{c_1} \cdot \bar \chi{}_{t_1}^{d_1}\right) \cdots
\left(\bar \chi{}_{s_p}^{c_p} \cdot \bar
\chi{}_{t_p}^{d_p}\right)\,,\quad p = n - 1\,.}  We have included
the dependence of this interaction on $v$ and $\bar v$. This
dependence can either be checked directly, or it can be deduced
from requirement that the interaction transforms as $\lambda^{-n}$
under $M\to \lambda M$, $\bar M\to \bar\lambda \, \bar M$, as
discussed in section 3.

To probe for the presence of the $F$-term, we thus compute in the
instanton background the correlation function\foot{Because the
correlator includes external legs with massless propagators, the
fermions conjugate to those in the effective vertex appear.}
\eqn\COR{ \left\langle \bar \chi \cdot \bar \chi
\left(\chi^{s_1}_{c_1} \cdot \chi^{t_1}_{d_1}\right) \cdots
\left(\chi^{s_p}_{c_p} \cdot
\chi^{t_p}_{d_p}\right)\right\rangle\,.} This computation as usual
has two pieces: a one-loop integral over fluctuating modes in the
instanton background and an integral over zero modes.  Because the
instanton background is supersymmetric to leading order, the
one-loop integral over quantum fluctuations is trivial and
contributes only a factor of unity.  So the important integral to
consider is the integral over zero modes.

\medskip\noindent{\it Bosonic Zero Modes}\smallskip

As usual, in the instanton background we have eight bosonic zero
modes.  Four zero modes are associated to the collective coordinate
$x_0$ for the location of the instanton in $\BR^4$.  One zero mode is
associated to the scale $\rho$ of the instanton.  Finally, three zero
modes arise from global $SU(2)$ gauge transformations and are
associated to a collective coordinate $h$ on $SU(2)$.

\medskip\noindent{\it Fermionic Zero Modes}\smallskip

Much more important than the bosonic zero modes are the fermionic zero
modes.  We have already discussed the counting of these modes
generally, but now we review the details.

First, we have two gaugino zero modes which arise from the action
of the chiral supercharges $Q_\alpha$ and which take the form
\eqn\LAMSUSY{ \lambda^{SS \, A \, [\beta]}_\alpha \,=\, {{\rho^2
\, \sigma^{A \, \beta}{}_{\! \! \! \alpha}} \over {(\rho^2 +
x^2)^2}}\,.} Here $SS$ stands for global supersymmetry, $A$ labels
the adjoint representation of $SU(2)$, $\alpha$ is a spinor index,
and $\beta$ simply labels the two zero modes.  Since we will not
try to compute the absolute normalization of our interaction, we
have not bothered to normalize the zero modes.

Second, at leading order in $g^2$, we have an additional $2n+2$
fermion zero modes.  Two of these extra zero modes are gaugino zero
modes associated to the action of the superconformal generators
$x^{\dot \beta}_\beta Q^\beta$, of the form
\eqn\LAMSC{ \lambda^{SC\, A \, [\dot \beta]}_\alpha \,=\, {{\rho \, x^{\dot
\beta}_\beta \, \sigma^{A \, \beta}{}_{\! \! \! \alpha}} \over {(\rho^2 +
x^2)^2}}\,.}

The other $2n$ zero modes arise from the $2n$ fermion doublets and
are of the form \eqn\PSIS{ \psi^i_{\alpha a [j]} \,=\, {{\rho \,
\delta^i_j \, h^b_a \, \epsilon_{\alpha b}} \over {(\rho^2 +
x^2)^{3/2}}} \,.} Again, $j$ is just an index that labels the zero
modes.  We have also included explicitly the dependence of these
modes on the element $h^b_a$ of $SU(2)$ parametrizing global gauge
transformations.  We could also have included this collective
coordinate in \LAMSUSY\ and \LAMSC, but any dependence of the
gaugino zero modes on $h$ will drop out immediately in our
computation.

These $2n$ zero modes transform in the representation ${\bf 2n}$
of the flavor group $SU(2n)$.  After giving expectations to the
quark superfields, $SU(2n)$ is broken to $SU(2)\times SU(2n-2)$
(where in an instanton field, $SU(2)$ must be combined with a
rotation).  Under the subgroup, the zero modes of $\psi$ transform
as $({\bf 2},{\bf 1})\oplus ({\bf 1},{\bf 2n-2})$.  The
superconformal zero modes similarly transform as $({\bf 2},{\bf
1})$.

\medskip\noindent{\it Yukawa Interactions}\smallskip

The zero modes in \LAMSUSY, \LAMSC, and \PSIS\ are simply zero
modes of the $\Dsl$ operator in the instanton background.
However, to perform the instanton computation, we must go beyond
leading order and consider the effect of the Yukawa couplings in
SQCD.  These couplings of course take the form \eqn\YUKAWA{ \int
\! d^4 x \; \bar q^a_i \, \left( \psi^i_b \cdot \lambda^b_a
\right) \,.}

On the Higgs branch, with $q$ satisfying \INSTB, this interaction
pairs the two superconformal zero modes $\lambda^{SC}$ with the
two  zero modes of the quarks that transform the same way, which
are those with $i=1,2$ in \PSIS\ (and which we have denoted $\chi$
in \COMPPHI). As a result, when we compute the correlator \COR,
these fermion zero-modes can be absorbed by pulling down two
copies of the Yukawa interaction \YUKAWA\ from the SQCD action,
which contributes a factor proportional to $\bar v^2$ to the
correlator.

We are then left with the two gaugino zero modes $\lambda^{SS}$
and the other $2n-2$ quark zero modes appearing in \PSIS.  Of
course, these $2n-2$ quark zero modes are absorbed directly by the
massless fermions $\chi^s_c$ appearing in the correlator \COR. But
what of the zero modes $\lambda^{SS}$?

To answer this question, we recall that another very important,
qualitative effect of the Yukawa coupling \YUKAWA\ is that it alters
the form of the zero modes $\lambda^{SS}$ to include components also
involving the fermion $\bar\chi$.  Specifically, to
first order in $\rho v$, the relevant equations of motion are
\eqn\CLASXLAM{ \Dsl \, \lambda \,=\, 0\,,\qquad \bar{\Dsl} \,\bar\psi
\,=\, \sqrt{2} \, \bar q \cdot \lambda\,,}
which have solution
\eqn\SOLNLAM{ \lambda \,=\, \lambda^{SS}\,,\qquad \bar \psi_{\dot
\alpha \, a}^{i \, [\beta]}  \,=\, {1 \over {4\pi}} \, \bar {\Dsl}_{\dot
\alpha}^{[\beta]} \bar q^i_a\,,}
with $\bar q$ as in \INSTB.  Simply by symmetry, the massless
components of $\bar \psi_{\dot \alpha \, a}^{i \, [\beta]}$ which mix
with $\lambda^{SS}$ must correspond to the singlet $\bar\chi$.  Thus,
the two supersymmetric zero modes $\lambda^{SS}$ are absorbed by the
two fermions $\bar \chi$ which appear in the correlator \COR.

The classical wavefunction of $\bar\chi$ can be explicitly evaluated in
the instanton background from \SOLNLAM, and far from the instanton
location $x_0$ the wavefunction takes the form
\eqn\CLASWV{ \bar\chi_{\dot \alpha}^{[\beta]}(x) \,=\, \bar v \rho^2
S^\beta_{\dot \alpha}(x, x_0)\,,}
where $S^\beta_{\dot\alpha}(x,x_0)$ is the free fermion propagator.

\medskip\noindent{\it Computing the Correlator}\smallskip

We are now prepared to compute the fermion correlator \COR\ in the
instanton background.  Using the classical wavefunctions \PSIS\
and \CLASWV\ for the fermion zero modes, we see that
\eqn\CORII{\eqalign{ &\left\langle \bar \chi \cdot \bar \chi
\left(\chi^{s_1}_{c_1} \cdot \chi^{t_1}_{d_1}\right) \cdots
\left(\chi^{s_p}_{c_p} \cdot
\chi^{t_p}_{d_p}\right)\right\rangle\,=\,\cr\noalign{\vskip 3 pt}
&\bar v^4 \, \Lambda^{6 - n} \, \int \! d^4 x_0 \, d \rho \, d \mu
\, \; \rho^{2n+5} \, \exp(- 4 \pi^2  \rho^2 |v|^2 / g^2) \,
\epsilon^{s_1 t_1 \cdots s_p t_p} \, \left( h_{c_1}^{e_1}
h_{d_1}^{f_1} \, \epsilon_{e_1 f_1} \right) \cdots
\left(h_{c_p}^{e_p} h_{d_p}^{f_p} \, \epsilon_{e_p f_p}
\right)\times\cr &\times \; \big(S(y_1 - x_0) \cdot S(y_2 -
x_0)\big) \cdots \big(S(y_{2n-1} - x_0) \cdot S(y_{2n} - x_0)
\big)\,.\cr}}

In this expression, $y_1,\ldots,y_{2n}$ are the positions of the
$2n$ fermions in $\BR^4$, which are assumed to be far from the
position $x_0$ of the instanton.  We then make use of the fact
that, in this limit, the classical wavefunctions \PSIS\ of the
fermions $\chi^s_c$ have the correct asymptotic behavior so that
the correlator can be written using the free fermion propagator
$S$. In computing the amputated vertex, we would simply drop these
factors and the integration over the position $x_0$ of the
instanton.

Besides the factor $d^4 x_0$, the bosonic measure also includes
a factor $d\rho \, \rho^{2n+5}$ and a factor $d\mu$, which represents the
invariant Haar measure on $SU(2)$.  We have determined the power of
$\rho$ that appears simply by dimensional analysis.

Thus, since a prefactor of $\bar v^4$ appears from the fermion zero
modes, the Gaussian integral over $\rho$ then produces the correct
dependence on $v$ and $\bar v$ as in \VRTX.  We have not been careful
about factors of the gauge coupling $g^2$ which also appear in the
integration measure and upon performing the Gaussian integral.  By
holomorphy, any explicit dependence of the correlator on $g^2$ should be
absorbed into a wavefunction renormalization of the external legs.

The only integral left to consider is the group integral over
$SU(2)$, which takes the form \eqn\SUTWO{ I^{d_1 \, d_2 \cdots \,
d_{2p}}_{c_1 \, c_2 \cdots \, c_{2p}} \,=\, \int \! d\mu \; h_{c_1}^{d_1}
\, h_{c_2}^{d_2} \cdots  h^{d_{2p}}_{c_{2p}}\,.} This integral is
not identically zero, since if one picks a real unit vector
$n^c_d$ and contracts with $n^{c_1}_{d_1}n^{c_2}_{d_2}\dots
n^{c_{2p}}_{d_{2p}}$, the integrand on the right hand side becomes
positive definite.  The $SU(2)\times SU(2)$ symmetry implies that
\eqn\BIGI{ I^{d_1 \, d_2 \cdots \, d_{2p}}_{c_1 \, c_2 \cdots \, c_{2p}}
\,\propto\, \epsilon^{d_1 d_2} \, \epsilon_{c_1 c_2} \cdots
\epsilon^{d_{2p-1} d_{2p}} \, \epsilon_{c_{2p-1} c_{2p}} \,+\,
(permutations)\,.} Here the first term on the right hand side must be
symmetrized under the exchanges of indices corresponding to exchanges
between the factors of $h$ in \SUTWO. (Note that, by bose statistics, a term
proportional to $\epsilon^{d_i d_j}$ for some $i$ and $j$ is also
proportional to $\epsilon_{c_i c_j}$.)  These symmetries arise in
the effective interaction \VRTX\ from the permutation symmetries
of the fermions.  Thus, upon substituting \BIGI\ into \CORII, we
produce the effective interaction which arises from the
multi-fermion $F$-term.

\subsec{A Computation in the Seiberg Dual With Six Doublets}

In many examples of duality, nonperturbative effects in the direct
theory become classical effects in the dual theory.  An example is
the $SU(2)$ gauge theory with $2n=4$, that is with four doublets.
In this case, the basic nonperturbative effect, as we reviewed in
section 2, is the deformation of the moduli space ${\cal M}$ of
vacua. The dual theory that describes the infrared physics is the
sigma model whose target is ${\cal M}$, and the complex structure
is built in at tree level.

Here we want to describe a more subtle example of this, for the
$SU(2)$ theory with $2n=6$ doublets.  In this case we will show
that the multi-fermion $F$-term that we have obtained
nonperturbatively in the direct description of SQCD can also be
computed at tree level in the Seiberg dual
\refs{\SeibergBZ,\SeibergPQ} description.

As promised in section 2, we also reconsider here the deformation
of complex structure that occurs in the theory with four doublets.
In particular, we reproduce the effective interaction in \DSEFFIV\
by integrating out the massive fields in the linear sigma model with
superpotential $W = \Sigma (M \^ M - \Lambda^4)$ which describes the
deformation.  Since this computation is exactly the same in spirit as our
classical computation in the Seiberg dual of the theory with six
doublets, we describe both computations together.

The Seiberg dual of $SU(2)$ SQCD with six doublets is
distinguished by the fact that the dual gauge group is trivial,
and hence this theory is especially simple.  In particular, the
elementary degrees of freedom in the dual theory are described
entirely by the mesonic fields $M^{ij}$, with Wess-Zumino action
\eqn\WZS{ S \,=\, {1 \over {\mu^2}} \, \int \! d^4 x \, d^4 \theta
\; \bar M M \,+\, \int \! d^4 x \, d^2 \theta \; \Lambda^{-3} \, M
\^ M \^ M \,+\, c.c.} We have included the canonical kinetic terms
in $S$, with an arbitrary scale $\mu$ that appears so that, by
convention, $M$ has dimension two.  Using a different kinetic term
for $M$ would not affect the computation of $F$-terms.

The cubic superpotential plays an interesting role in this theory.
As shown by Seiberg \SeibergBZ, this potential appears
nonperturbatively in the electric theory, but in the dual theory
it arises at tree level.  In either case, the $F$-term
equations which follow from this superpotential are simply the
classical Pl\"ucker relations $M \^ M = 0$ that enforce the
condition $\rk{M} \le 2$, which is necessary to describe ${\cal
M}$.

In the special case $n = 3$, the multi-fermion $F$-term in
\COMIV\ takes the explicit form \eqn\WEFF{\eqalign{ \delta S \,=\,
{1 \over {\mu^4}} \, \int \! d^4 x \, &d^2 \theta \; \Lambda^3 \,
(\bar M M)^{-3} \, \epsilon^{i_1 j_1 i_2 j_2 i_3 j_3} \, \bar
M_{i_1 j_1} \, \times\cr &\times\,\left(M^{kl} \, \bar D \mskip 2 mu \bar
M_{i_2 k} \cdot \bar D \mskip 2 mu \bar M_{l j_2}\right) \, \left( M^{k'l'} \,
\bar D \mskip 2 mu \bar M_{i_3 k'} \cdot \bar D \mskip 2 mu \bar M_{l' j_3}
\right)\,.\cr}} We will generate this effective interaction in the
most naive way possible.  We simply observe that, when we expand
the Wess-Zumino model around a generic point on ${\cal M}$, the
cubic superpotential induces a mass for some components of $M$.
We then integrate out these massive modes at tree level in a
Feynman diagram computation to generate \WEFF.

At this point, one might immediately protest that we are making
the quixotic proposal to generate an $F$-term in perturbation
theory and in blatant violation of standard non-renormalization
theorems.  However, these non-renormalization theorems have only
been considered for conventional $F$-terms which describe
superpotentials, and the multi-fermion $F$-terms we consider evade
them in an interesting way.

The essential point here is that the multi-fermion $F$-terms arise
from cohomology classes on ${\cal M}$.  Whenever we perform a
perturbative computation around some vacuum on ${\cal M}$, we are
only working in a small neighborhood of that point, and in that
neighborhood any operator $\CO_\omega$ which represents a positive
degree cohomology class of $\bar Q{}_{\dot \alpha}$ becomes $\bar
Q{}_{\dot \alpha}$-trivial.  As a result, though globally on
${\cal M}$ the multi-fermion $F$-terms cannot be written as
$D$-terms, they can be written as $D$-terms if we expand in
fluctuations around a given vacuum.  These $D$-terms can then be
directly generated in perturbation theory.

As a simple and highly relevant example, we consider the $F$-term
at hand in \WEFF.  We expand \WEFF\ around some point with
$\langle  M^{ij} \rangle \neq 0$.  With no loss of generality, we
can assume that the only nonzero component of $\langle
M^{ij}\rangle$ is $\langle M^{12}\rangle$.  In expanding around
this particular vacuum, we apply our standard convention that $c,d,e,f$ refer
to indices $1,2$, $s,t,u,v$ refer to indices $3,\ldots,6$, and
$i,j,k,l$ run over all indices $1,\ldots,6$.  From \WEFF, we generate
a series of interactions among the fluctuating fields $\delta M$, one
interaction being
\eqn\WEFFII{\eqalign{ \delta S \,=\, {1 \over {\mu^4}} \, \int \!
d^4 x\, &d^2 \theta \; \Lambda^3 \, \langle \bar M M \rangle^{-3}
\,  \, \langle \bar M_{12} \rangle \, \times\cr
&\times\,\epsilon^{s_1 t_1 s_2 t_2 }\left( \delta M^{c d} \, \bar
D \, \delta \bar M_{s_1 c} \cdot \bar D \, \delta \bar M_{d t_1}
\right) \, \left( \delta M^{e f} \, \bar D \, \delta \bar M_{s_2
e} \cdot \bar D \, \delta \bar M_{f t_2} \right)\,.\cr}} Of
course, the effective fermion interaction \VRTX\ which we
considered in the instanton computation is one of the terms that
arises from \WEFFII.

By definition, if $\delta \bar M_{i j}$ is massless, then the
basic equation of motion \EOMII\ for $\delta \bar M_{i j}$ takes
the form $\bar D^2 \, \delta \bar M_{i j} \,=\, \CO(\delta M^2)$.
Since only massless fluctuations appear in the effective
interaction \WEFFII, we can immediately integrate this $F$-term
into a $D$-term at leading order, \eqn\WEDD{\eqalign{ \delta S
\,=\, {1 \over {\mu^4}} \, \int \! d^4 x \, &d^4 \theta \;
\Lambda^3 \, \langle \bar M M \rangle^{-3} \, \, \langle \bar
M_{12} \rangle \, \times\cr &\times\, \epsilon^{s_1 t_1 s_2 t_2 }
\left( \delta M^{c d} \, \delta \bar M_{s_1 c} \, \delta \bar M_{d
t_1}\right) \, \left( \delta M^{e f} \, \bar D \, \delta \bar
M_{s_2 e} \cdot \bar D \, \delta \bar M_{f t_2} \right)\,.\cr}} We
have used the fact that to this order, two $\bar D$'s cannot act
on the same $\delta\bar M$, and none can act on $\delta M$.

In the case of the theory with $n = 2$, the same observations
imply that the analogous part of the $F$-term in \DSEFFIV\ can be
rewritten locally as the simple $D$-term below, \eqn\WEDDII{
\delta S \,=\, {1 \over {\mu^2}} \int \! d^4 x \, d^4 \theta \;
\Lambda^4 \, \langle \bar M M \rangle^{-1} \,
\epsilon^{c d}\epsilon^{s t} \, \delta \bar M_{s c} \delta \bar
M_{d t}\,.} Here again we expand around a vacuum in which the
nonvanishing part of $\langle M\rangle $ is $\langle
M^{12}\rangle$, and $c,d=1,2$ while $s,t=3,4$.

Thus, the appearance of these unusual $F$-terms is signaled by the
perturbative appearance of the $D$-terms in \WEDD\ and \WEDDII,
which we must now compute.  As in the instanton computation, we
could compute some particular component of this superspace
interaction.  However, we are in a situation perfectly suited for
a manifestly supersymmetric computation using the formalism of
super Feynman diagrams.

\medskip\noindent{\it Evaluating a Super Feynman Diagram}\smallskip

We will not review here the basic derivation of Feynman rules in
superspace, for which we recommend section 6.3 of \GatesNR.  In
general, superspace Feynman rules can be derived by standard path
integral manipulations just as for ordinary Feynman rules, and for
the sake of brevity we will only state the super Feynman rules
that we need for our very simple, tree-level computations.

In the case of the theory with $2n=6$, we begin by expanding the
tree-level Wess-Zumino action in fluctuations $\delta M$ about the
vacuum, so that \eqn\WZSII{  S \,=\, {1 \over {\mu^2}} \, \int \!
d^4 x \, d^4 \theta \; \delta \bar M \delta M \,+\, \int \! d^4 x
\, d^2 \theta \left( 3 \, \lambda \, \langle M \rangle \^ \delta M
\^ \delta M \,+\, \lambda \, \delta M \^ \delta M \^ \delta M
\right)+\, c.c.,} where for convenience we introduce the
abbreviation \eqn\COUPL{ \lambda \,\equiv\, \Lambda^{-3}\,.} We
will not be concerned with constants here, and we simply absorb
the numerical factor of $3$ in \WZSII\ into $\langle M \rangle$.
We will also suppress the appearance of the mass scale $\mu$ in
all expressions that follow, since its appearance is trivially
fixed at the end of the computation by dimensional analysis.

Of course, we similarly expand the sigma model action in the
theory with $n = 2$, \eqn\WZSIII{ S \,=\, {1 \over {\mu^2}} \,
\int \! d^4 x \, d^4 \theta \; \delta \bar M \delta M \,+\, \int
\! d^4 x \, d^2 \theta \left( 2 \, \langle M \rangle \^ \delta M
\, \delta\Sigma \,+\, \delta M \^ \delta M \, \delta \Sigma \,-\,
\varepsilon \, \delta \Sigma \right)+\, c.c. \,+\, \cdots\,,}
where the ellipses indicate kinetic terms and a mass term for the
fluctuations of the auxiliary field $\Sigma$.  As above, we ignore
constants, and we abbreviate \eqn\COUPLII{ \varepsilon \,\equiv\,
\Lambda^4\,.} The most important terms in \WZSIII\ for our
computation are simply the linear source term for $\delta \Sigma$
which represents the deformation as well as the mass term mixing
$\delta M$ and $\delta \Sigma$.

\medskip\noindent{\it Propagators}\smallskip

In the vacuum with only $\langle M^{1 2}\rangle\not= 0$, we want to
get an effective interaction for the massless fields by
integrating out the massive fields $M^{s t}$, $s,t=3,\dots,2n$.

These fields have  standard superspace propagators, which may be
either chiral or non-chiral.  We indicate these propagators below,
in the theory with $n = 3$, \eqn\PROP{\matrix{\noalign{\vskip 5pt}
&\delta \bar M_{s t} \hbox{
--------------- } \delta M^{u v} &= &\delta^{u v}_{s t}\, /
\left(p^2 + \bar \lambda \lambda \langle \bar M M
\rangle\right)\,,\cr\cr &\delta M^{s t} \buildrel{D^2}\over{\hbox{
-- -- -- -- -- -- }} \delta M^{u v} &= &\bar \lambda \, \langle
\bar M_{12} \rangle \, \epsilon^{ s t u v} \, D^2 \,/\, p^2
\left(p^2 + \bar \lambda \lambda \langle \bar M M
\rangle\right)\,,\cr\cr &\delta \bar M_{st} \buildrel{\bar
D^2}\over{\hbox{ -- -- -- -- -- -- }} \delta \bar M_{u v} &=
&\lambda \, \langle M^{12} \rangle \, \epsilon_{ s t u v} \, \bar
D^2 \,/\, p^2 \left(p^2 + \bar \lambda \lambda \langle \bar M M
\rangle\right)\,.\cr}} In writing the non-chiral propagator, we
use the standard notation $\delta^{u v}_{s
t}=\delta^u_s\delta^v_t-\delta^u_t\delta^v_s$. We have also
suppressed a superspace delta function $\delta^4(\theta -
\theta')$ which accompanies these propagators. Finally, we note
the superspace derivatives $D^2$ and $\bar D^2$ which appear in
the chiral and anti-chiral propagators. These factors arise
ubiquitously in supergraph computations when chiral integrals over
half of superspace are rewritten as non-chiral integrals over the
full superspace.

In the theory with $n = 2$, similar propagators appear for the
appropriate linear combinations of $\delta \Sigma$ and $\delta M$,
for which the mass squared is again proportional\foot{If a
separate mass term $m \Sigma^2$ for $\Sigma$ is also present, this
statement remains true in the classical limit that $\langle \bar M
M \rangle$ is large.} to $\langle \bar M M \rangle$.

\medskip\noindent{\it Vertices}\smallskip

In the theory with $n = 3$, the cubic superpotential gives rise to
cubic vertices for chiral and anti-chiral interactions, as we
distinguish in Figure 1.  We have written these interactions in an
$SU(6)$ symmetric fashion, though of course each chiral and
anti-chiral vertex decomposes under the unbroken $SU(2) \times SU(4)$
symmetry to give various interactions between the massive and massless
components of $M$, which we leave implicit.  Each superspace vertex
comes with a factor of $\int d^4 \theta$, and the delta functions from
the propagators simply ensure that the overall diagram has precisely
one factor of $\int d^4 \theta$, as we expect. \midinsert
$$\matrix{
&\lower 23pt\hbox{{\epsfxsize=1in\epsfbox{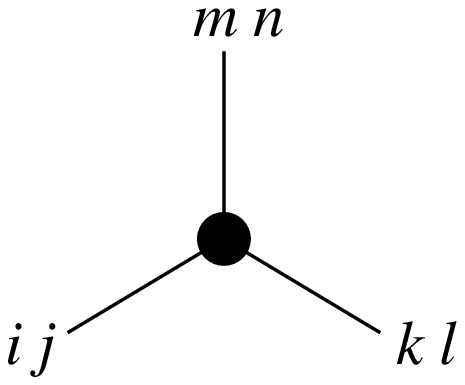}}}&=&\lambda \,
\epsilon_{i j k l m n}\,,\cr\noalign{\vskip 10 pt}
&\lower 23pt\hbox{{\epsfxsize=1in\epsfbox{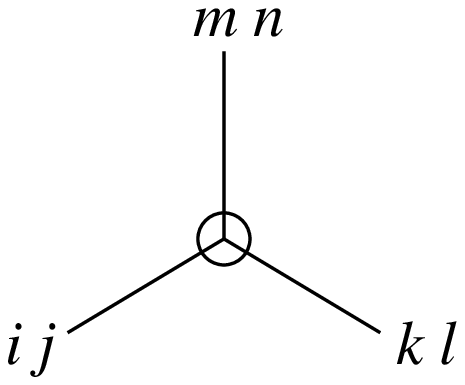}}}&=&\bar \lambda
\, \epsilon^{i j k l m n}\,.\cr}$$
\smallskip\centerline{\it Figure 1.  Vertices for $n = 3$}\smallskip
\endinsert\noindent

In the corresponding theory with $n = 2$, we require a similar
cubic vertex arising from the interaction $\delta \bar M \^ \delta
\bar M \, \delta \bar \Sigma$ as well as the chiral source term
$\varepsilon \, \delta \Sigma$, as shown in Figure 2. \midinsert
$$\matrix{
&\lower 23pt\hbox{{\epsfxsize=1in\epsfbox{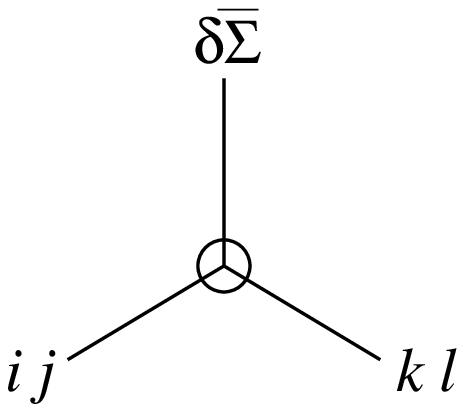}}}&=&\epsilon^{i j
k l}\,,\cr\noalign{\vskip 10 pt}
&\lower 8 pt\hbox{{\epsfxsize=1in\epsfbox{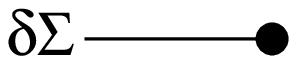}}}&=&\varepsilon\,.\cr}$$
\smallskip\centerline{\it Figure 2.  Vertices for $n = 2$}\smallskip
\endinsert\noindent
Again, we leave the obvious decomposition under $SU(2) \times SU(2)$
implicit.

Last, we recall the rule that if a chiral vertex has $N$ internal
legs (external legs don't count), then $N-1$ of those legs appear
with a factor of $\bar D^2$ attached.  Briefly, if $J(x,\theta)$
is the chiral source introduced as usual to derive Feynman rules,
then the functional derivative of $J$ satisfies $\delta
J(x,\theta) / \delta J(x',\theta') = \bar D{}^2 \, \delta^4(x-x')
\, \delta^4(\theta - \theta')$. So $N$ factors of $\bar D{}^2$
appear from these derivatives, but one factor of $\bar D{}^2$ is
used to write $\int d^2 \theta \, \bar D{}^2 = \int d^4 \theta$,
as mentioned above.

With these rules in hand, we can immediately generate the
interactions in \WEDD\ and \WEDDII.  First, in the simpler case of
$n = 2$, we immediately evaluate the simple diagram in Figure 3 at
zero momentum to produce the effective interaction \eqn\FDII{ \int
\! d^4 x \, d^4 \theta \; \epsilon^{c d} \epsilon^{s t} \, \delta \bar M_{s
c} \delta \bar M_{d t} \, {\varepsilon \over {\langle \bar M M
\rangle}}\,,} as in \WEDDII. \midinsert
\centerline{\epsfxsize=2in\epsfbox{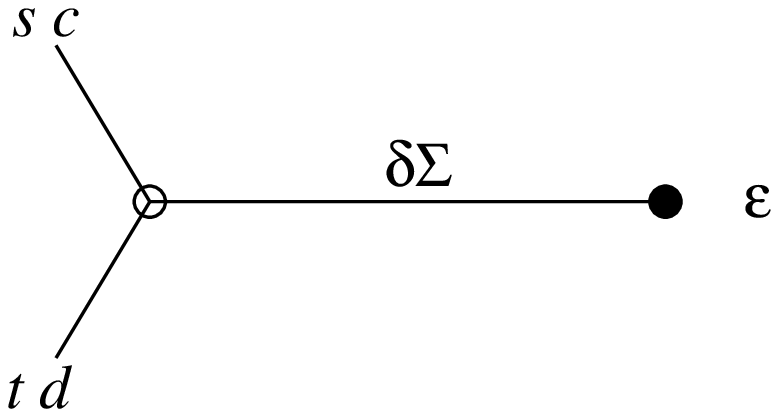}}
\medskip\centerline{\it Figure 3.  Two-point super Feynman diagram}\medskip
\endinsert

For the theory with $n = 3$, we consider the slightly more
involved diagram in Figure 4.  We note that the $D^2$ operator in
this diagram arises from the central chiral propagator, and the
two $\bar D{}^2$ operators arise from the two chiral vertices.
\midinsert \centerline{\epsfxsize=4in\epsfbox{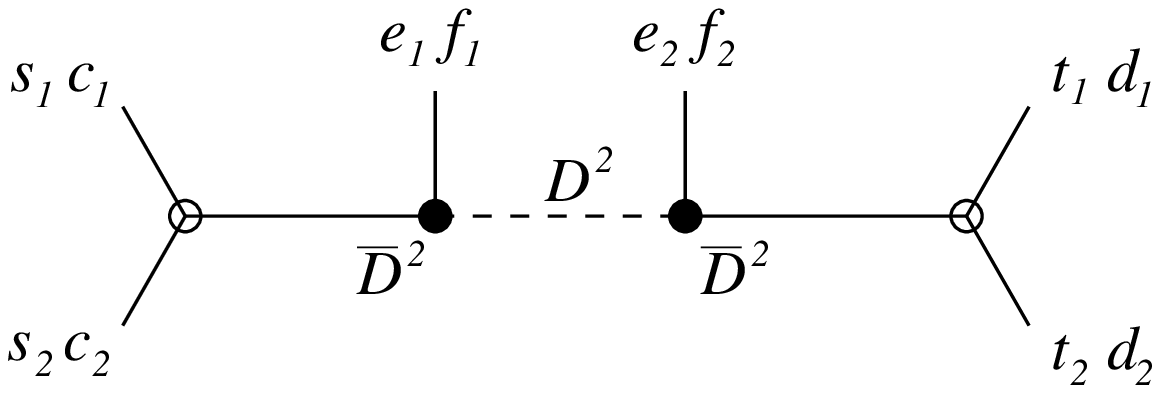}}
\medskip\centerline{\it Figure 4.  Six-point super Feynman diagram}\medskip
\endinsert

At first sight, one might worry about the spurious pole at zero momentum
that appears to arise from the extra factor of $p^2$ appearing in the
central chiral propagator, as in \PROP.  Physically, since we only
integrate out massive fields, we do not expect to find any pole at zero
momentum.

However, we can integrate by parts to move one of the $\bar D{}^2$
operators onto the central chiral propagator to form $\bar D{}^2
D^2$.  Since $\bar D{}^2 D^2 = p^2$ when acting on a chiral field,
this factor of $\bar D{}^2 D^2$ cancels against the extra factor
of $p^2$ in the denominator of the chiral propagator.  Thus, the
diagram is well defined in the limit of zero momentum, and we
evaluate it in this limit to reproduce the $D$-term \WEDD.  We
also note that once we cancel the factor of $\bar D{}^2 D^2$, we
are left with only one factor of $\bar D{}^2$, which acts on the
external anti-chiral legs just as in the interaction \WEDD.

So at zero momentum, the remainder of our computation is a trivial
matter of algebra.  We find that this diagram produces the
effective interaction \eqn\FD{\eqalign{ \int \! d^4 x \, &d^4
\theta \; \delta \bar M_{s_1 c_1} \, \delta \bar M_{s_2 c_2} \,
\left(\bar D\delta \bar M_{t_1 d_1} \cdot \bar D \delta \bar
M_{t_2 d_2}\right) \, \delta M^{e_1 f_1} \, \delta M^{e_2
f_2}\times\cr &\times {{\bar\lambda \, \epsilon^{c_1 c_2}
\epsilon^{s_1 s_2 u v}} \over {\bar \lambda \lambda \, \langle
\bar M M \rangle}} \cdot \lambda \, \epsilon_{u v u' v'}
\epsilon_{e_1 f_1} \cdot {{\bar \lambda \, \epsilon^{u' v' w x}
\langle \bar M_{1 2} \rangle} \over {\bar \lambda \lambda \,
\langle \bar M M \rangle}} \cdot \lambda \, \epsilon_{w x w' x'}
\epsilon_{e_2 f_2} \cdot {{\bar \lambda \, \epsilon^{w' x' t_1
t_2} \epsilon^{d_1 d_2}} \over {\bar \lambda \lambda \, \langle
\bar M M \rangle}}\,.\cr}}  The tensor on the second line of \FD\
is then proportional to \eqn\FDII{ \lambda^{-1} \, \langle \bar M
M \rangle^{-3} \, \langle \bar M_{1 2} \rangle \; \delta^{c_1
c_2}_{e_1 f_1} \; \delta^{d_1 d_2}_{e_2 f_2} \; \epsilon^{s_1 s_2 t_1
t_2}\,,} which has precisely the form required to produce the
$F$-term. The $\bar\lambda$'s have happily canceled, ensuring the
requisite holomorphy.

N. Seiberg pointed out the following interpretation of the
$1/\lambda$ factor.  As the meson superfield $M$ has dimension two
in the classical theory, the superpotential interaction
proportional to $\int \! d^4x \, d^2\theta \, M\wedge M\wedge M$
must on dimensional grounds be interpreted in the  underlying SQCD
theory as $\int \! d^4x \, d^2\theta \, \Lambda^{-3} \, M\wedge
M\wedge M$. Thus, what we have called $\lambda$ is a multiple of
$\Lambda^{-3}$ in SQCD, as in \COUPL.  Hence the
 multi-fermion $F$-term  interaction, being proportional to
 $\lambda^{-1}$ in the Seiberg dual description, is proportional
to $\Lambda^3$ in the original SQCD description.  $\Lambda^3$ is the
standard instanton factor for $SU(2)$ with six doublets, and the
direct instanton computation of section 4.1 did, accordingly, give a
result proportional to $\Lambda^3$.

\subsec{Mass Deformation And Renormalization Group Flow}

For our final computation, we perturb $SU(2)$ SQCD with $2 n$
massless doublets by adding a tree-level superpotential which
gives a mass to some of the $n$ flavors, \eqn\TREEW{ W \,=\, m_{i
j} \, M^{i j}\,.} As usual, we assign charges to the mass
parameters $m_{i j}$ under the symmetries of the massless theory
so that $W$ is formally invariant, \eqn\CHRGM{\matrix{
&\quad&SU(2)\quad&SU(2n)\quad&U(1)_A\quad&U(1)_\CR\quad\cr
\noalign{\vskip 2 pt} &m_{ij}\quad&{\bf
1}\quad&\bigwedge^2\!\left({\bf\bar{2n}}\right)\quad&-2\quad&-2\left({1
- {2 \over n}}\right)\,.\quad\cr}} The whole computation will be
performed on $B$.

As we  observed in general in section 2.3, the tree-level
superpotential alters the on-shell supersymmetry algebra of the
theory. Consequently, the operator $\CO_\omega \equiv
\CO_\omega^{(n)}$ in \COMIV\ which is chiral in SQCD with $2n$
massless doublets is no longer chiral when some of those doublets
become massive.

Physically, we expect that there is instead some deformation
$\wt\CO_\omega$ of this operator, depending holomorphically on $m_{i
j}$, which is chiral in the massive theory and which reduces to
$\CO_\omega^{(n)}$ upon setting $m_{i j}$ to zero.

On the other hand, if we give very large masses to $k$ of the
flavors and integrate them out, we also expect that
$\wt\CO_\omega$ must reduce to the operator $\CO_\omega^{(n-k)}$
appropriate for the massless theory with $n-k$ flavors.  In
particular, upon integrating out all but one flavor,
$\wt\CO_\omega$ should reproduce the well known nonperturbative
superpotential, \eqn\NPW{ W = {{\Lambda^5} \over {M}}\,.} Here
$M=M^{12}$ is the only independent component of the $2\times 2$
antisymmetric matrix $M^{ij}$.

We now compute $\wt\CO_\omega$, which will be uniquely determined
from $\CO_\omega^{(n)}$ by supersymmetry and will have the
properties above. Since we know already from the work of \AffleckMK\
that the superpotential \NPW\ is generated, we will thus show that the
$F$-term involving $\CO_\omega^{(n)}$ is generated in the massless
theory with $n$ flavors.  Finally, we remark that this sort of
analysis extends, at least in spirit, directly to the general case
of $SU(N_c)$ SQCD with $N_f > N_c$ flavors and might be
successfully applied there.

As before, we use ${\cal M}$ to denote the moduli space of the
massless theory, and we recall that $\CM$ is a complex cone over the
Grassmannian $B = SU(2n) / S(U(2) \times U(2n-2))$.  Then our problem
of constructing $\wt\CO_\omega$ is equivalent to the geometric problem
of finding a tensor $\wt\omega$, which is generally an inhomogeneous
sum of sections of $\bar\Omega^p_{\cal M} \otimes \bigwedge^p T{\cal M}$ for
various $p$, such that $\wt\omega$ satisfies  the supersymmetry
condition, \eqn\WTOM{ \left(\bar\partial + \iota_{dW}\right) \wt\omega \,=\,
0\,,} and in the massless limit reduces to our former tensor $\omega$.

\medskip\noindent{\it Preliminaries}\smallskip

As in section 3, the important analysis of $\wt\omega$ is the local
analysis on $B$ near the point corresponding to $M^{i j} =
\hat\epsilon^{i j}$.  However, we first find it useful
to revisit our construction of the simpler tensor $\omega$ in greater
detail and in a manner which immediately generalizes to the construction
of $\wt\omega$.

Let us  recall our construction of $\omega$ in section 3.  We
begin by picking a point $P$ on $B$, for concreteness
corresponding to the point $M^{i j} = \hat\epsilon^{i j}$ on
$\CM$. Since the $M^{ij}$ are homogeneous coordinates on $B$, we
can use the scaling symmetry to set $M^{12}=1$.  Then we take
complex coordinates on $B$ (in a neighborhood of the point $P$) to
be simply the off-diagonal matrix elements
$\phi^s_c=\epsilon_{cd}M^{sd}$, $c=1,2$, $s=3,\dots,2n$. They
transform as in \SYMSII\ under the action of the unbroken $S(U(2)
\times U(2n-2))$ symmetry group at $P$. The matrix elements
$M^{ij}$, $i,j>2$, are determined in terms of the $\phi^s_c$ by
the equation $M\wedge M=0$.  We will not need the explicit form of
these matrix elements; the only important fact is that they are of
order $(\phi^s_c)^2$.

In \COMIV, we determined the form of the multi-fermion $F$-term:
\eqn\jurrto{ \CO_\omega \,=\, \Lambda^{6-n} \, \left(\bar M
M\right)^{-n} \, \epsilon^{i_1 j_1 \cdots i_n j_n} \, \bar M_{i_1
j_1} \, \CO_{i_2 j_2} \cdots \CO_{i_n j_n},} where \eqn\nmurg{
\CO_{ij} \,\equiv\, M^{kl} \, \bar D \,\bar M_{ik} \cdot \bar D
\,\bar M_{lj}\,,\qquad \bar M M = \ha \bar M_{ij} M^{ij}\,.}
 In that
discussion, we used an argument based on symmetries to prove that
$\bar\partial\omega=0$.  As a prelude to including the
superpotential deformation, we will here demonstrate this more
explicitly.

Since $\omega$ is invariant under the action of $SU(2n)$ on the
homogeneous space $B$, it suffices to show that
$\bar\partial\omega=0$ at the point $P$.  As $\bar\partial$ is a
first order differential operator, to evaluate
$\bar\partial\omega$ at $P$, it suffices to describe $\omega$ near
$P$ to within terms of order $\phi^2$.  This leads to drastic
simplification.  For example, $\bar M M = 1 +{\cal O}(\phi^2)$, so
we can simply set $\bar MM=1$.  Furthermore, examination of
\jurrto\ and \nmurg\ shows that up to terms of order $\phi^2$, we
can replace the explicit factor of $\bar M_{i_1j_1}$ in \jurrto\
by $\hat\epsilon_{i_1j_1}$, so that all factors of ${\cal
O}_{i_kj_k}$ have $i_k,j_k>2$.  Further, for $i_k,j_k>2$, we can
take ${\cal O}_{i_kj_k}=\epsilon^{cd}{\bar D \, \bar M_{i_k c}}\cdot
{\bar D}\, {\bar M}_{d j_k}$, again up to terms of order
$\phi^2$. The net effect is that, up $\phi^2$ terms,
\eqn\trufr{\omega=\epsilon^{s_1t_1s_2t_2\dots
s_nt_n}\left(\epsilon_{c_1d_1}d\bar\phi{}_{s_1}^{c_1}{\partial\over\partial
\phi^{t_1}_{d_1}}\right)\cdots\left(\epsilon_{c_nd_n}
d\bar\phi{}_{s_n}^{c_n}{\partial\over\partial
\phi^{t_n}_{d_n}}\right).} Now the fact that
$\bar\partial\omega=0$ at $P$ is completely obvious: all terms in
$\omega$ have constant coefficients (up to terms of order $\phi^2$
that have been dropped), so there is nothing that the derivatives
in $\bar\partial$ can act on.

The benefit of this approach is that we can now conveniently
understand the generalization with the superpotential turned on.
We claim that the generalization of ${\cal O}_\omega$ is simply
\eqn\noz{\eqalign{ \wt \CO_\omega \,&=\, \Lambda^{6-n} \,
\left(\bar M M\right)^{-n} \, \epsilon^{i_1 j_1 \cdots i_n j_n} \,
\bar M_{i_1 j_1} \, \wt \CO_{i_2 j_2} \cdots \wt \CO_{i_n
j_n}\,,\cr \wt \CO_{ij} \,&\equiv\, M^{kl} \, \bar D \mskip 2
mu\bar M_{ik} \cdot \bar D \mskip 2 mu \bar M_{lj} -
\left(\bar M M\right) \, m_{i j}\,.\cr}} This certainly reduces to
${\cal O}_\omega$ at $m=0$; we just have to prove that it is
chiral.  In other words, we need to show that the object
$\tilde\omega$, obtained from $\omega$ by replacing each
$\CO_{ij}$ by $\wt\CO_{ij}$, is annihilated by
$\bar\partial+\iota_{dW}$.  It suffices to do the computation at
the point $P\in B$ with $M^{ij}=\widehat \epsilon^{ij}$ since, as
we will make no particular assumption about the form of the mass
matrix $m_{ij}$, the computation would proceed in the same way at
any other point.

So as before, we want to write out a simple formula for
$\tilde\omega$ that is valid near $P$ to within an error of order
$\phi^2$.  The same reasoning applies as before: the explicit
factors of $\bar M_{i_1j_1}$ in $\tilde{\cal O}_\omega$ and of
$M^{kl}$ in $\tilde{\cal O}_{ij}$ can be replaced by
$\hat\epsilon_{i_1j_1}$ and $\hat\epsilon^{kl}$.  Since in
$\tilde{\cal O}_{ij}$, the indices $i,j$ are then in the range
$3,\dots,2n$ the mass matrix $m_{ij}$ can be replaced by
$\mu_{ij}$, its orthogonal projection onto the part with $i,j>2$.
We write $\Pi$ for the projector onto components of $m$ with
$i,j>2$ and will describe $\Pi$ more explicitly momentarily.

The upshot is that up to terms of order $\phi^2$, $\wt\omega$ is
described near $P$ by a simple generalization of \trufr,
\eqn\WTLCOM{\eqalign{ \wt\omega \,&=\, \epsilon^{s_1 t_1 \cdots
s_p t_p} \,  \wt\omega_{s_1 t_1} \cdots \wt\omega_{s_p
t_p}\,,\quad p=n-1\,,\cr \wt\omega_{s t}
&=\epsilon_{cd}\left(d\bar\phi{}_s^c{\partial\over
\partial\phi^t_d}+{\partial\over\partial\phi^s_c}d\bar\phi{}_t^d\right)
- \mu_{s t}\,.}} The virtue of factorizing $\wt\omega$ in this way
is as we will see each factor $\tilde\omega_{s_it_i}$ is
separately annihilated by $\bar\partial+\iota_{dW}$.  Also, in the
expression for $\wt\omega_{s t}$ in the second line of \WTLCOM, we
have explicitly indicated the two terms that arise from the
contraction of spinor indices on $\bar D_{\dot \alpha}$ in \noz,
since we will try to be careful about factors of two in the
following.

Let us first evaluate $\iota_{dW}(\tilde \omega_{st})$.  The
contraction operator $\iota_{dW}$ trivially annihilates $\mu_{st}$
(because the latter is a zero-form).  As $W=m_{ij}M^{ij}$, we have
$dW=m_{ij}dM^{ij}$.  So the effect of contraction with $dW$ is
just to map $\partial/\partial\phi^s_c$ to $\mu^c_s$, the
projection of the mass matrix $m$ to terms that transform like
$\partial/\partial \phi^s_c$ (in other words, as $({\bf
2},\bar{\bf 2n-2})$) under the subgroup of the symmetry group that
leaves fixed the point $P\in B$.  Hence we have \eqn\CONTRACTII{
\iota_{dW} \wt\omega_{s t} \,=\, \epsilon_{c d} \left(
d\bar\phi{}^c_s \, \mu^{d}_t \,+\, \mu^c_s \,
d\bar\phi{}^d_t\right)\,.}

It remains to evaluate $\bar\partial(-\mu_{st})$.  This is nonzero
because of the projection in the definition of $\mu_{st}$. As we
will show, \eqn\DMIJ{ \bar\partial \mu_{s t} \,=\, \epsilon_{c d}
\left(d\bar\phi{}^c_s \, \mu_t^d \,+\, \mu_s^c \,
d\bar\phi{}^d_t\right)\,.} From \CONTRACTII\ and \DMIJ, we then see
directly that $\bar\partial + \iota_{dW}$ annihilates
$\wt\omega_{s t}$ and hence $\wt\omega$ at the point $P$ on $B$.

To derive the formula \DMIJ\ for $\bar\partial \mu_{s t}$, we
begin by considering the projection $\Pi$ of the mass matrix $m$
onto its components which transform in the representation
$\wedge^2\left(\bar{\bf 2n-2}\right)$. We can directly write a
global formula for this projection, \eqn\PROJM{\eqalign{ \Pi(m)_{i
j} \,&=\, m_{i j} \,+\, \left(\bar M M\right)^{-1} \left( m_{i
k} M^{k l} \bar M_{l j} - m_{j k} M^{k l} \bar M_{l i}\right)
\,+\,\cr &+\, \left(\bar M M\right)^{-2} \left(\bar M_{i k} M^{k
l} m_{l p} M^{p q} \bar M_{q j}\right)\,.\cr}} Upon substituting
$M^{i j} = \hat \epsilon^{i j}$ and using repeatedly that $\hat
\epsilon^{k l} \hat \epsilon_{l j} = - \hat \delta^k_j$
(explaining the signs above), one can check that the second and
third terms of \PROJM\ subtract the components of $m$ transforming
in the representations ${\bf 1}$ and $({\bf 2},\bar{\bf 2n
- 2})$ under $SU(2) \times SU(2n-2)$ at $P$, leaving only the
components in $\wedge^2\left(\bar{\bf 2n-2}\right)$.  Since the
formula \PROJM\ for $\Pi$ is invariant, it is correct globally on $B$.

Because the action of $\bar\partial$ commutes with pullback, we
can now act with $\bar\partial$ directly on \PROJM\ as an
unconstrained expression in the ambient vector space (or
projective space) parametrized by $M^{i j}$.  We then pull this
expression back to $\CM$ by dropping all terms which involve the
one-forms $d \bar M_{i j}$ with both indices $i,j > 2$.

To evaluate $\bar\partial\mu$ at $\phi=0$, we can discard all
terms proportional to $\phi$, and in particular to components
$M^{ij}$ or $\bar M_{ij}$ with $i$ or $j$ bigger than 2.  Terms
that survive at $\phi=0$ only arise from the action of
$\bar\partial$ on the second term of \PROJM, with the expression
\eqn\PROJMII{ \left( \bar M M \right)^{-1} \left( m_{i k} M^{k l}
d \bar M_{l j} - m_{j k} M^{k l} d\bar M_{l i}\right)\,, \quad i,j
> 2\,.}  From this global expression \PROJMII\ we immediately
deduce the local formula \DMIJ\ upon setting $M^{i j} =
\hat\epsilon^{i j}$ and identifying $m_{i k} M^{k l} d\bar M_{l
j}$ as representing locally $\epsilon_{c d} \, \mu^c_s \,
d\bar\phi{}^d_t$ at $P$.  We remark that the relative sign between
the two terms in \DMIJ\ and \PROJMII\ arises  from a rearrangement
of flavor indices in passing from \PROJMII\ to \DMIJ.

Finally, although we have thus far only considered the special
case that $W = m_{i j} M^{i j}$, if we now consider the case of a
general superpotential deformation of SQCD, then our construction
of $\wt\omega$ immediately generalizes upon substituting
everywhere $\partial W / \partial M^{i j}$ for $m_{i j}$.  The
only important property of $m$ which we used was the fact that it
is annihilated by $\bar\partial$, which is always true for $dW$.

\bigskip\noindent{\it Renormalization Group Flow}

To conclude, we consider how $\wt\CO_\omega$ in \noz\ behaves under
renormalization group flow.  If we expand $\wt\CO_\omega$ as a
polynomial in $m$, then the term of degree $k$ in $m$ is given by
\eqn\COZVII{\eqalign{
\CO_\omega^{(n-k)} \,&=\, (-1)^k \, {{n-1} \choose k} \;
\Lambda^{6-n} \, \left(\bar M M\right)^{-(n-k)} \,\epsilon^{i_1 j_1
\cdots i_n j_n} \times \cr
&\times \, \bar M_{i_1 j_1} \, m_{i_2 j_2} \cdots m_{i_{k+1}
j_{k+1}} \,\CO_{i_{k+2} j_{k+2}} \cdots \CO_{i_n j_n}\,,\cr
\CO_{ij} \,&\equiv\, M^{kl} \, \bar D \,\bar M_{ik} \cdot \bar D \,\bar M_{lj}\,.\cr}}
This operator $\CO_\omega^{(n-k)}$ has the same form as the operator
in \COMIV\ which appears in the theory with $n-k$ massless flavors.

We consider the limit in which $k$ flavors have masses $m \gg
\Lambda$.  To integrate out these flavors, we restrict to the
sublocus of $\CM$ describing supersymmetric vacua in the massive
theory, so that $m_{i k} M^{k j} = 0$ for all $i,j$ (as follows
from the $F$-term equations), and we simply omit from the operator
$\wt\CO_\omega$  any terms which involve the heavy quarks. The
operator to which $\wt\CO_\omega$ flows in the infrared is thus
$\CO_\omega^{(n-k)}$ in \COZVII.

In particular, we can consider flowing to the theory with only one
flavor.  The operator to which $\wt\CO_\omega$ flows is then given
by \eqn\CONE{ \CO_\omega^{(1)} \,=\, (-1)^{(n-1)}\, \Lambda^{6-n}
\, \left(\bar M M\right)^{-1} \,\epsilon^{i_1 j_1 \cdots i_n j_n}
\, \bar M_{i_1 j_1} \, m_{i_2 j_2} \cdots m_{i_n j_n}\,.} As we
see, $\CO_\omega^{(1)}$ involves no fermions at all and represents
a function on ${\cal M}$.  Of course, this function is not
holomorphic on all of ${\cal M}$.

However, if we restrict $\CO_\omega^{(1)}$ to the sublocus of
${\cal M}$ describing supersymmetric vacua, then
$\CO_\omega^{(1)}$ is holomorphic.  Indeed, this locus can be
described by a single massless meson $M$, so the matrix structure
disappears and $\bar M$ cancels out.  On this locus,
$\CO_\omega^{(1)}$ can be written in terms of  $M$ as \eqn\NPWII{
\CO_\omega^{(1)} \,=\, (-1)^{(n-1)} \, \Lambda^{6-n}
\, \epsilon^{i_1 j_1 \cdots i_{n-1} j_{n-1}} \, m_{i_1 j_1} \cdots
m_{i_{n-1} j_{n-1}} \, {2 \over M}\,.} In this expression, the
Pfaffian of the rank $2(n-1)$ minor of $m$ appears, and an extra
factor of two arises from the contraction of indices of $\bar M_{i_1
j_1}$.  So, once ultraviolet and infrared scales are matched,
$\CO_\omega^{(1)}$ reproduces the nonperturbative superpotential in \NPW.

\bigbreak\bigskip\bigskip\centerline{{\bf Acknowledgements}}\nobreak

We would like to thank G. Moore and N. Seiberg for helpful
suggestions on this work.  CB would also like to thank A. Dymarsky,
S. Murthy, P. Ouyang, D. Shih, and P. Svr\v cek for stimulating
conversations.  Finally, CB would like to thank the participants of
the 2004 Simons Workshop, where a preliminary version of this work was
presented, for numerous comments and suggestions.

The work of CB is supported in part under National Science Foundation Grants
No. PHY-0243680 and PHY-0140311, and the work of EW is supported in part under
National Science Foundation Grant No. PHY-0070928.  Any opinions,
findings, and conclusions or recommendations expressed in this
material are those of the authors and do not necessarily reflect
the views of the National Science Foundation.

\listrefs

\end